\documentclass[%
 reprint,
superscriptaddress,
 aip,
 amsmath,amssymb,
]{revtex4-1}

\usepackage{dcolumn}
\usepackage{bm}

\usepackage{graphicx}

\begin{document}

\preprint{AIP/123-QED}

\title{From time-series to complex networks: application to the cerebrovascular flow patterns in atrial fibrillation}

\author{Stefania Scarsoglio} 
\email{stefania.scarsoglio@polito.it} 
\affiliation{Department of Mechanical and Aerospace Engineering, Politecnico di Torino, Torino, Italy}
\author{Fabio Cazzato}%
\affiliation{Medacta International SA, Castel San Pietro, Switzerland}
\author{Luca Ridolfi}
\affiliation{Department of Environmental, Land and Infrastructure Engineering, Politecnico di Torino, Torino, Italy}%

\begin{abstract}
A network-based approach is presented to investigate the cerebrovascular flow patterns during atrial fibrillation (AF) with respect to normal sinus rhythm (NSR). AF, the most common cardiac arrhythmia with faster and irregular beating, has been recently and independently associated with the increased risk of dementia. However, the underlying hemodynamic mechanisms relating the two pathologies remain mainly undetermined so far; thus the contribution of modeling and refined statistical tools is valuable. Pressure and flow rate temporal series in NSR and AF are here evaluated along representative cerebral sites (from carotid arteries to capillary brain circulation), exploiting reliable artificially built signals recently obtained from an \textit{in silico} approach. The complex network analysis evidences, in a synthetic and original way, a dramatic signal variation towards the distal/capillary cerebral regions during AF, which has no counterpart in NSR conditions. At the large artery level, networks obtained from both AF and NSR hemodynamic signals exhibit elongated and chained features, which are typical of pseudo-periodic series. These aspects are almost completely lost towards the microcirculation during AF, where the networks are topologically more circular and present random-like characteristics. As a consequence, all the physiological phenomena at microcerebral level ruled by periodicity - such as regular perfusion, mean pressure per beat, and average nutrient supply at cellular level - can be strongly compromised, since the AF hemodynamic signals assume irregular behaviour and random-like features. Through a powerful approach which is complementary to the classical statistical tools, the present findings further strengthen the potential link between AF hemodynamic and cognitive decline.
\end{abstract}

\maketitle

\begin{quotation}
The paper presents a network-based perspective to investigate the cerebrovascular flow patterns during atrial fibrillation (AF) with respect to normal sinus rhythm (NSR). There has been recently growing evidence that AF, the most common cardiac arrhythmia with faster and irregular beating, is independently associated to the increased risk of dementia. The topic has a high social impact given the number of individuals involved and the expected increasing AF incidence in the next forty years. Although several mechanisms try to explain the relation between the two pathologies, causality implications are far from being clear. In particular, little is known about the distal and capillary cerebral circulation. Thus, the contribution of modeling and statistical tools is valuable. Here, we exploit the powerful techniques of complex network theory to better characterize the cerebrovascular patterns (in terms of \textit{in silico} pressure and flow rate time-series) during AF with respect to NSR. Each cerebral region (from large carotid arteries to capillary districts) is represented through a network, whose topological features evidence the differences between arrhythmic and normal beating, thus highlighting in the micro-circulation plausible mechanisms for cognitive decline during AF.
\end{quotation}

\section{Introduction}

The growing size of numerical data coming from multiscale simulations, highly-resolved imaging and computational fluid dynamics approaches \cite{Etchings} requires refined quantitative tools to appropriately analyze biomedical signals. Complex network theory, by combining elements from the graph theory and statistical physics, offers an innovative and synthetic framework to handle and interpret complex systems with a huge number of interacting elements \cite{albert_barabasi_2002,Boccaletti_et_al_2006,WS1998}. In the last decades, beside the well-established applications to Internet, World Wide Web, economy and social dynamics \cite{costa_et_al_2011,Havlin}, complex networks have been employed in a variety of physical and engineering systems, e.g. from hydrology and climate dynamics (e.g., \cite{Sivakumar,Donges_et_al_2009,scarsoglio2013,Tsonis_Swanson_2008}), turbulent and fluid flows (e.g., \cite{Taira,Scarsoglio_IJBC_2016,Tupikina}), to biomedical applications (e.g., \cite{Reijneveld,Lusis,Avila,Bullmore}). The network-based approach has been extensively proposed for the characterization of time series which - by means of algorithms based on recurrence plot \cite{Donner}, visibility graph \cite{Lacasa}, correlation matrix \cite{Donner} and pseudo-periodicity \cite{Zhang_2006} - are converted into complex networks.

We here present a network-based approach, which relies on the correlation matrix and exploits the pseudoperiodic framework proposed for the electrophysiological signals \cite{Zhang_2006,Zhang_et_al_2008}, to study the cerebrovascular flow patterns during atrial fibrillation (AF). This cardiac pathology, characterized by irregular and accelerated heart-beating, is the most common arrhythmia with an estimated number of 33.5 million individuals affected worldwide in 2010 \cite{Chugh_2014}. Beside the well-known disabling symptoms (such as palpitations, chest discomfort, anxiety, decrease of blood pressure, limited exercise tolerance, pulmonary congestion) which deteriorate the quality of life, AF is related to thromboembolic transient ischemic attacks \cite{Buchwald_2016} and increases the risk of suffering an ischemic stroke by five times \cite{Wolf_1991}. Through a collection of hemodynamic mechanisms, such as silent cerebral infarctions \cite{Kalantarian_2014,Gaita_2013}, altered cerebral blood flow \cite{Sabatini_2000}, hypoperfusion \cite{Jacobs_2014} and microbleeds, an independent association between AF and cognitive decline has been recently observed.

Although an increasing number of very different observational studies suggests this possible association (e.g., among the most recent \cite{Kalantarian_2013,Hui_2015,Chen_2016,Thacker_2013}), to the best of our knowledge none of them definitely states a causal relation between AF hemodynamics and cognitive impairment. Recent works highlight AF consequences on the cerebral circulation (e.g. lower diastolic cerebral perfusion and decreased blood flow in the intracranial arteries), but the causal connections still remain mostly undetermined \cite{Jacobs_2014,Hui_2015,Kanmanthareddy}. Moreover, current techniques - such as transcranial doppler ultrasonography and intracranial  pressure measures - are difficult to be obtained and usually fail in capturing the cerebral micro-vasculature fluid dynamics. For all these reasons, little is known about AF effects on altered pressure levels and irregular cerebral blood flow. However, an accurate cerebral hemodynamic mapping would be helpful in revealing AF feedbacks on brain circulation. For example, anomalous pressure levels could be symptomatic to predict haemorrhagic events such as microbleeds, while impair blood flow repartition can locally alter brain oxygenation.

Awaiting direct clinical evidences which are nowadays lacking, the efficiency of the computational hemodynamics is a promising branch \cite{Severi,Shi}, as it can be extremely helpful in isolating single cause-effect relations and understanding which AF features lead to specific hemodynamic changes. In recent works, AF and normal sinus rhythm (NSR) cerebral hemodynamic signals were \textit{in silico} simulated and compared \cite{SR_2016,Scarsoglio_JRSI}, revealing a dramatically altered scenario during AF, where critical events - such as hypoperfusions and hypertensions - are more likely to occur in the peripheral brain circulation.

The aim of the present work is to exploit a validated modeling algorithm to analyze, through some suitable advanced metrics adopted in the complex network theory, how the peripheral cerebral circulation is altered by AF, in particular, in those regions where \textit{in vivo} measures are still missing. Hemodynamic signals of pressure and flow rate over 1000 heartbeats in NSR and AF are evaluated along four representative cerebral sites (internal carotid artery, middle cerebral artery, distal region, capillary-venous district), exploiting the \textit{in silico} data recently obtained \cite{SR_2016}. A total number of sixteen time series is thus analyzed. Each time series is then transformed into a network, associating with each temporal segment a node and identifying possible links among nodes through the linear correlation coefficient. If the statistical interdependence between two nodes is above a sufficiently high threshold, a link between the two nodes exists. The resulting sixteen networks are then studied with the network metrics, revealing in the peripheral circulation different topological features between NSR and AF.

\section{Methods}

A stochastic modeling approach is adopted to reproduce pressure and flow rate signals in the cerebral circulation during NSR and AF. Two lumped-parameter models describing the cardiovascular and cerebral circulations are exploited in series and stochastically forced through artificially-built RR series, where RR (s) is the temporal range between two consecutive heart beats. The cardiovascular and cerebral models have been previously calibrated and extensively validated as long as literature data are available \cite{Korakianitis,MBEC_2014,CMBBE_2016,SR_2016,Ursino_2010} to mimic realistic NSR and AF conditions for the cerebrovascular circulation. Thus, the validated \textit{in silico} outputs are the object of the present work and used to understand how the cerebral fluid dynamics and the temporal structure of the hemodynamic signals modify during AF, especially in the peripheral regions where clinical measures are still lacking. This goal is achieved through a refined and innovative method for signal analysis, based on the complex network theory, by broadening and strengthening what recently observed with more classical statistical tools \cite{Scarsoglio_JRSI}.

\subsection{Cerebral time series: pressures and flow rates in NSR and AF}

\begin{figure*}
\includegraphics[width=1.61\columnwidth]{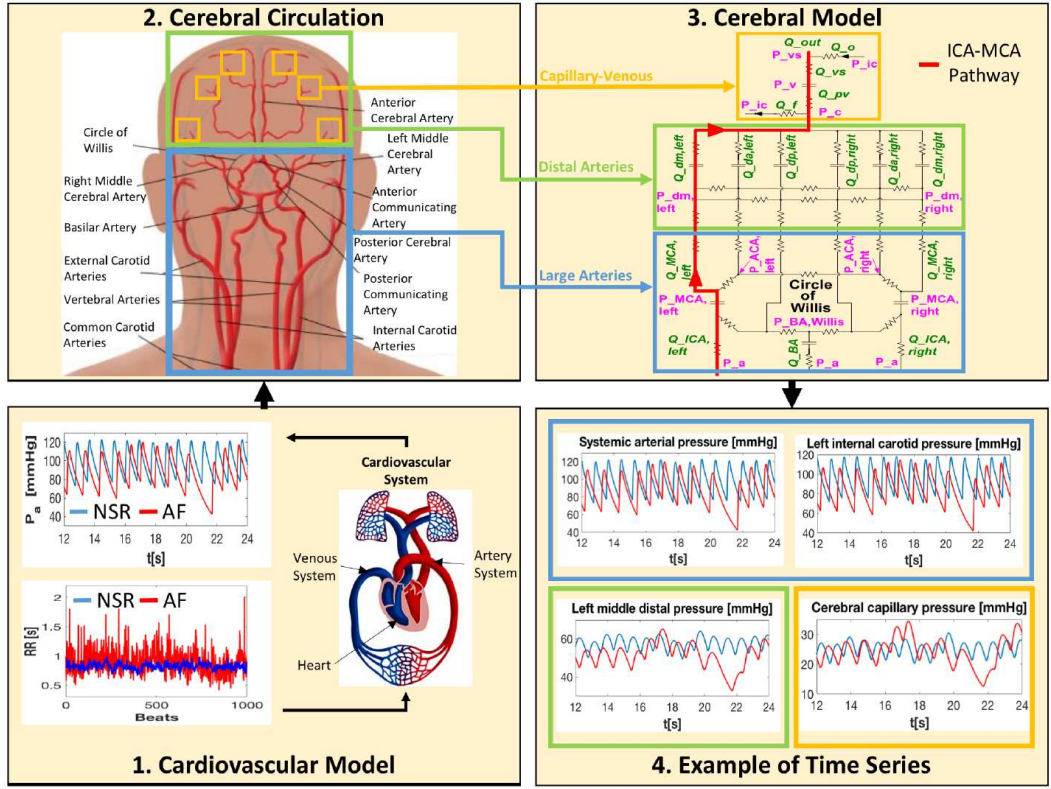}
\caption{Schematic algorithm of the modeling approach. (1) Cardiovascular model. 1000 extracted RR intervals in NSR (blue) and AF (red), and examples of $P_a$ time series obtained through the cardiovascular model. (2) Cerebral circulation. Scheme of the cerebral vasculature forced by the $P_a$ signal. (3) Cerebral model.  $R$: resistance, $C$: compliance, $Q$: flow rate, $P$: pressure. The left ICA-MCA pathway is evidenced in red and is composed by $P_a$, $Q_{ICA,left}$, $P_{MCA,left}$, $Q_{MCA,left}$, $P_{dm,left}$, $Q_{dm,left}$, $P_c$ and $Q_{pv}$. (4) Example of time series. Representative pressure time series along the ICA-MCA pathway, in NSR (blue) and AF (red) conditions.}
\end{figure*}

The NSR and AF signals have been acquired from a recent computational hemodynamic study \cite{SR_2016}, which consists of an algorithm made of three subsequential steps. In Figure 1, the modeling approach adopted (panels 1, 2, 3) is described and representative pressure time series obtained as outputs (panel 4) are displayed. The present computational framework combines two different lumped models in sequence: the cardiovascular model is run to obtain the systemic arterial pressure, $P_a$, which is next exploited as forcing input for the cerebral model.

To obtain the cerebral time series, the first step is to artificially generate the RR intervals, in NSR and AF conditions. We used artificially-built RR-intervals to avoid the patient-specific details (e.g., sex, age, weight, cardiovascular diseases, ...) inherited by real RR beating. To purely catch the overall impact of a fibrillated cardiac status, we consider the same healthy young adult configuration, first forced through NSR and then by AF rhythm. Being normal beating a remarkable example of pink noise \cite{Hayano_1997}, RR beats during NSR have been extracted from a pink-correlated Gaussian distribution (mean $\mu=0.8$ s, standard deviation, $\sigma=0.06$ s), which is the typical distribution observed during sinus rhythm \cite{MBEC_2014,Hennig_2006}. Differently to the white noise which is uncorrelated, the adoption of a pink noise introduces a temporal correlation, which is a common feature of the normal beating \cite{Musha1982,Peng1993,Saul1987,Yamamoto1994}. The AF distribution is commonly unimodal (60-65\% of the cases) and is fully described \cite{Hennig_2006,Hayano_1997} by the  superposition  of  two   statistically  independent  times  ,  $RR=\varphi+\eta$.  $\varphi$  is  extracted  from  a  correlated  pink  Gaussian  distribution,   $\eta$  is instead  drawn  from  an uncorrelated exponential distribution (rate parameter $\gamma$). The resulting AF beats are thus obtained from an uncorrelated Exponentially Gaussian Modified distribution (mean $\mu=0.8$ s, standard deviation $\sigma=0.19$ s, rate parameter $\gamma=7.24$ Hz). The AF beating results less correlated and with a higher variability than NSR, as clinically observed \cite{Hennig_2006,Bootsma,Hayano_1997}. The RR parameters of both configurations are  suggested  by  the available data \cite{Hennig_2006,Sosnowski,Hayano_1997}, and by considering that the coefficient of variation, $cv$, is around 0.24 during AF \cite{Tateno}. Since these RR time sequences have been validated and tested over clinically measured beating \cite{Hennig_2006,Hayano_1997}, we adopt them as the most suitable and reliable RR time-series to model NSR and AF conditions. Moreover, both RR series have been chosen with the same mean heart rate (75 bpm) to facilitate comparison between the two conditions. More details on RR extraction are reported elsewhere \cite{MBEC_2014}.

\noindent 1000 cardiac cycles, shown in Fig. 1 (first panel), are considered for each configuration. This allows us to achieve statistically stable results. Namely, we tested that other extractions of the same number of beats give results analogous to those described in the following sections III and IV.

As second step, a lumped cardiovascular model was run using the RR signals as input, to obtain the signals of systemic arterial pressure, $P_a$, in NSR and AF conditions (Fig. 1, first panel), which will be the forcing inputs of the cerebral model. The cardiovascular modeling, first proposed by Korakianitis and Shi \cite{Korakianitis} and then widely adopted in the computational hemodynamics \cite{Shi}, consists of a network of electrical components - such as compliances, resistances and inductances - and describes the whole systemic and pulmonary circulation, with an active representation of the four heart chambers. Parameters were fitted to reproduce the physiological hemodynamics of a young healthy man, by providing results which are in agreement with the expected real behavior \cite{westerhof1}. Moreover, by suitably changing the parameters, the computational approach is able to capture the main features of different cardiovascular pathologies, such as hypertension and valvular diseases \cite{Korakianitis,PeerJ_2016}. During AF, the model was first tuned and extensively validated in resting conditions over more than thirty clinical studies \cite{MBEC_2014,CMBBE_2016}. Then, it has been successfully adopted to evaluate different AF aspects, such as the concomitant presence of left valvular diseases \cite{PeerJ_2016} and the role of increased heart rate at rest \cite{PlosOne_2015} and under effort \cite{PlosOne_2017}. To mimic the absence of the atrial kick, which is observed in AF, left and right atria are imposed as passive.

In the end, the $P_a$ signals are introduced in a lumped parameter model which simulates the entire cerebral circulation \cite{Ursino_2010} (Fig.1, second panel). The cerebral model is based on electrical counterparts and accurately describes the cerebral circulation up to the peripheral and capillary regions. It is able to reproduce several different pathological conditions characterized by heterogeneity in cerebrovascular hemodynamics and can be divided into three main sections: large arteries (Fig. 1, light blue box), distal arterial circulation (Fig. 1, green box), and capillary/venous circulation (Fig. 1, yellow box). The computational approach has been validated mainly over mean flow rates up to the middle cerebral circulation \cite{SR_2016}, since current clinical techniques for pressure and flow rate measures lack the resolving power to give any insights on the micro-vasculature. For this reason, the cerebral modeling can be a useful predictive tool to understand how the hemodynamic signals change towards the micro-circulation in presence of AF \cite{SR_2016,Scarsoglio_JRSI}. The left vascular pathway ICA-MCA (i.e., internal carotid artery - middle cerebral artery) evidenced in Fig. 1 (Panel 3, red path) is here analyzed as representative of the blood flow and pressure distributions from large arteries to the capillary-venous circulation: left internal carotid artery ($P_a$ and $Q_{ICA,left}$), middle cerebral artery ($P_{MCA,left}$ and $Q_{MCA,left}$), middle distal district ($P_{dm,left}$ and $Q_{dm,left}$), capillary-venous circulation ($P_c$ and $Q_{pv}$). Pressure time series of the left ICA-MCA pathway are reported as exemplificative of NSR and AF behaviors, in the fourth panel of Fig. 1. More details on the cerebral model are offered elsewhere \cite{SR_2016}. All the cerebral pressure and flow rate time series selected to build the corresponding networks have a frequency of 250 Hz and are composed, as previously mentioned, by 1000 heartbeats.

\subsection{Complex network metrics}

Some fundamental concepts of the complex network theory are here summarized \cite{albert_barabasi_2002,Boccaletti_et_al_2006}, by recalling only the measures and definitions which are relevant to the present analysis.

\noindent A network is defined by a set $V = 1,...,N$ of nodes and a set $E$ of links $\{i,j\}$. We assume that a single link exists between a pair of nodes and no self-loops can occur. The \emph{adjacency matrix}, $A$:

\begin{equation}
\label{adjacency} A_{ij} = \begin{cases} 0, & \mbox{if } \{i,j\} \notin E\\ 1, & \mbox{if } \{i,j\} \in E, \end{cases}
\end{equation}

\noindent accounts whether a link is active or not between nodes $i$ and $j$. The network is considered as undirected, thus $A$ is symmetric, moreover $A_{ii}=0$ since self-loops are not allowed.

\noindent The normalized degree centrality of the $i-th$ node is defined as

\begin{equation}
\label{dc} k_i = \frac{\sum\limits_{j=1}^N A_{ij}}{N-1},
\end{equation}

\noindent and represents the number of neighbors of node $i$, normalized over the total number of possible neighbors ($N-1$).

\noindent The \emph{eigenvector centrality}, measuring the influence of the node $i$ in the network, is given by

\begin{equation}
x_i = \frac{1}{\lambda} \sum_k A_{ki} x_k,
\end{equation}

\noindent where $A_{ki}$ is the adjacency matrix and $\lambda$ is its largest eigenvalue (in modulus) \cite{newman2010}. In matrix notation, we have:

\begin{equation}
{\lambda} x = x A,
\end{equation}

\noindent where the centrality vector $x$ is the left-hand eigenvector of the adjacency matrix $A$ related to the eigenvalue $\lambda$.

\noindent The \emph{assortativity coefficient}, $r$, \cite{Newman_2002} is defined as

\begin{equation}
r = \frac{1}{\sigma_q^2} \sum_{jk}jk (e_{jk} - q_j q_k),
\end{equation}

\noindent where $q_j$ is the distribution of the remaining degree, that is the number of edges leaving the node other than the one we arrived along. $e_{jk}$ is the joint probability distribution of the remaining degrees of the two nodes, $j$ and $k$ (for undirected networks, $e_{jk}=e_{kj}$). This quantity follows the rules: $\sum_{jk}e_{jk}=1$ and $\sum_j e_{jk}=q_k$. The assortativity coefficient is the Pearson correlation coefficient of the degree between pairs of linked nodes, thus $r \in [-1,1]$. When $r = 1$, the network has perfect assortative mixing patterns, meaning that high-degree nodes tend to connect each other (e.g., rich-club effect). When $r = 0$ the network is non-assortative or uncorrelated, which is typical of random graphs, while at $r = -1$ the network is completely disassortative, that is high-degree nodes are linked to low-degree nodes.

\noindent The \emph{local clustering coefficient} of node $i$ is

\begin{equation}
lc_i = \frac{e(\Gamma_i)}{\frac{k_i(k_i-1)}{2}},
\end{equation}

\noindent where $\Gamma_i$ is the set of first neighbors of $i$, $e(\Gamma_i)$ is the number of edges connecting the vertices within the neighborhood $\Gamma_i$, and $k_i(k_i-1)/2$ is the maximum number of edges in $\Gamma_i$, $0 \le lc_i \le 1$. The local clustering coefficient gives the probability that two randomly chosen neighbors of $i$ are also neighbors. The \emph{global clustering coefficient} is the mean value of $lc_i$, $\overline{lc} = \sum_{i=1}^N lc_i /N$.

\noindent The \emph{betweenness centrality} of node $k$ is

\begin{equation}
bc_k = \sum_{i,j \neq k} \frac{\sigma_{ij}(k)}{\sigma_{ij}},
\end{equation}

\noindent where $\sigma_{ij}$ are the number of shortest paths connecting nodes $i$ and $j$, while $\sigma_{ij}(k)$ represents the number of shortest paths from $i$ to $j$, across node $k$. If node $k$ is crossed by a large number of all existing shortest paths (i.e., high $bc_k$ values), then it can be considered an important mediator for the information transport in the network.

\noindent The \emph{closeness centrality} of node $i$ is

\begin{equation}
cc_i = \frac{N-1}{\sum\limits_{j=1}^N d_{ij}}
\end{equation}

\noindent where the shortest path length, $d_{ij}$, is the minimum number of edges that have to be crossed from node $i$ to node $j$, with $i, j \in V$ ($d_{ii}=0$). If $i$ and $j$ are not connected, the maximum topological distance in the graph $d_{ij} = N-1$ is used in the sum. Closeness centrality is normalized as follows, $0 \leq cc_i \leq 1$. According to this definition, node $i$ has a high closeness centrality value when it is topologically close to the rest of the network.

\subsection{Building the networks: from time-series to complex networks}

To transform the time series into complex networks, the approach proposed for pseudo-periodic series \cite{Zhang_2006,Zhang_et_al_2008} has been adopted. The complete temporal signal is divided into sequential cycles according to the RR intervals, which represent the portions of series corresponding to each beat. Fig. 2a shows an example of pressure temporal series divided into 5 time segments, according to the RR beating. Every temporal segment is associated to a node of the network, with the convention that node $i$ corresponds to the $i-th$ beat ($i\in [1,1000]$). Therefore, nodes are ordered in agreement with the beating sequence. In the example reported in Fig. 2, the network is thus composed by 5 nodes.

\begin{figure}
\includegraphics[width=0.81\columnwidth]{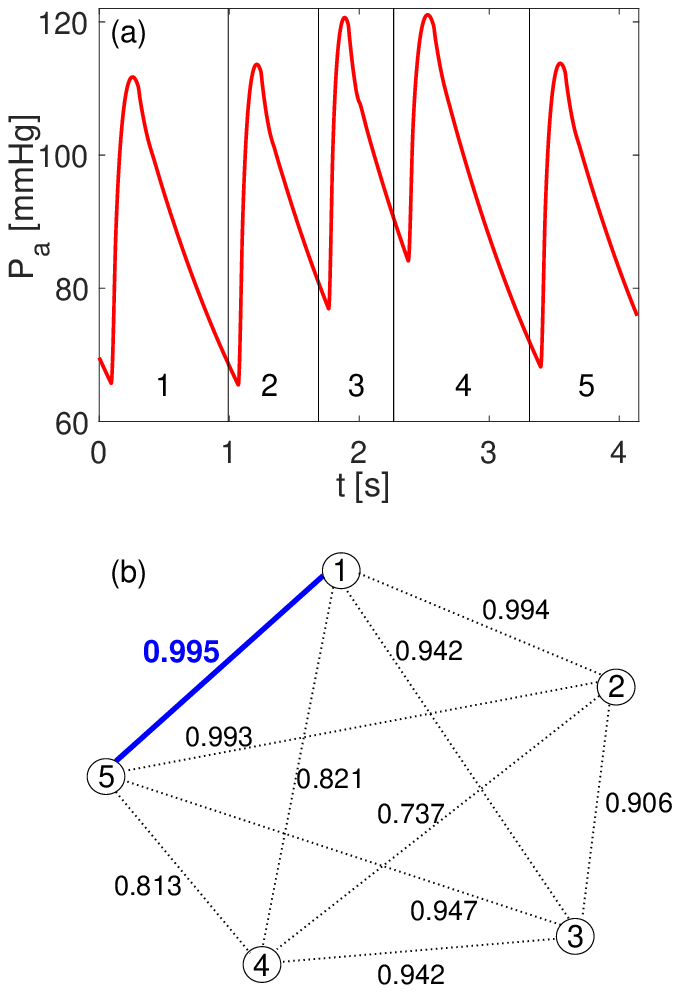}
\caption{(a) Example of pressure time series at the internal carotid entrance, $P_a$. Each $RR$ interval is transformed into a node. (b) Example of network built from the pressure time series reported in panel (a), by taking $R_{ij}\geq\tau$, where $\tau$ is the ninth decile of the matrix $R$.}
\end{figure}

\noindent For each hemodynamic signal the cross-correlation matrix $R$ is built, where the element $R_{ij}$ represents the maximum value of the linear correlation coefficient between the $i-th$ and $j-th$ temporal segments. The maximum correlation value is taken when the two segments have different temporal lengths, by shifting the shortest signaling segment all along the length of the longest segment. Each matrix $R$ is symmetric ($R_{ii}=1$ by definition) and has dimension $n$ x $n$, where here $n=1000$ is the number of the heartbeats (i.e., nodes) analyzed. Exploiting the symmetry of the $R$ matrix, the correlation coefficients to be evaluated are the matrix elements above the diagonal and correspond to the number of possible links, $n(n-1)/2$. In the example of Fig. 2b, the correlation matrix is (5x5), thus we have 10 correlation coefficients.

To compare pairs of cycles, the maximum value of the correlation coefficients has been selected among other distance criteria. Two other distance measures have been checked. The first analyzed is the phase space distance \cite{Zhang_et_al_2008}, defined as

\begin{equation}
M_{ij} = \min_{l} \frac{1}{\min(l_i,l_j)} \sum\limits_{k=1}^{\min(l_i,l_j)} ||X_k - Y_{k+1}||,
\end{equation}

\noindent where $l_i$ and $l_j$ are the lengths of the cycle $i$ and $j$, while $X_k$ and $Y_k$ are the $k-th$ elements (e.g., hemodynamic signals) of the cycles $i$ and $j$, respectively. $M$ represents the minimum value of the sum of the modules of the differences between the samples of all the possible pairs of cycles. The second measure is based on the mean value distance. For each cycle, $i$, a mean value of the hemodynamic signal is computed, $x_i$. Then, the distance matrix, $D$, is built considering the distance, $d_i$, between the mean value of the cycle $i$ and the average value of the complete signal. The element $D_{ij}$ is defined as $|d_i - d_j|$.

\noindent The three distance measures here introduced lead to similar results, however the most significative to evidence the difference between NSR and AF conditions turned out to be the correlation matrix, $R$. Therefore, results in Sections III and IV are presented only by using the linear correlation.

To define the adjacency matrix, $A$, and the corresponding network, a link between nodes $i$ and $j$ exists whether $R_{ij}\geq\tau$, where $\tau$ is the ninth decile of the matrix $R$ (computed excluding the diagonal). In so doing, each resulting network is undirected ($A_{ij}=A_{ji}$) and unweighted, since $A_{ij}=1$ (unitary weight) as long as $R_{ij}\geq\tau$. In the example of Fig. 2b, the ninth decile of the corresponding $R$ matrix is $0.995$, thus two nodes are connected by a link if the corresponding correlation coefficient is equal or above this value. The active link (1-5) is highlighted through a thick blue line.

\noindent The choice of the threshold, $\tau$, has been long discussed in correlation-based networks \cite{Donges_et_al_2009,Donner} and is a non-trivial aspect of building the network. It should represent a good compromise between a very high degree of correlation and a suitable network dimension. Our goal is to compare the networks here built, which represent different cerebral regions (from the internal carotid to the capillary regions) in physiological (NSR) and pathological (AF) conditions. Therefore, we preferred using a threshold which, case by case, through the ninth decile accounts for the maximum correlation of the local dynamics considered, rather than a unique threshold (e.g., $\tau=0.8$) fixed for all the configurations, which could be not equally meaningful in all the districts and conditions. Since $\tau$ is the only arbitrary parameter involved in the network building, a sensitivity analysis with different $\tau$ percentile values (namely, $85^{th}$ and $95^{th}$ percentiles) is reported in the Appendix A.

The described mapping of time series into networks has been achieved for pressure and flow rate series during NSR and AF in the 4 aforementioned cerebral regions. Thus, 16 networks are built and analyzed starting from the corresponding 16 hemodynamic series.

\section{Results}

Outcomes for the metrics introduced in Section IIB are here presented for the 16 networks. Apart from the assortativity which is a global network parameter, each node of a network has a value for the analyzed metrics (degree centrality, eigenvector centrality, local clustering coefficient, betweenness centrality and closeness centrality). To synthesize this large amount of information, we build the probability density function (PDF) and the cumulative function (CDF) distributions for all the metrics of each network. Then, at the same district and for the same hemodynamic signal (pressure and flow rate) we compare NSR and AF conditions. We recall that, given a beating condition (NSR or AF), results shown in the following are insensitive to the specific RR values composing the sequence of 1000 cardiac cycles, since this number of beats allows the statistical stationary to be reached. In fact, we checked that other extractions with the same number of beats give analogous trends, with negligible differences with respect to the differences observed between NSR and AF.

\begin{figure}
\includegraphics[width=0.81\columnwidth]{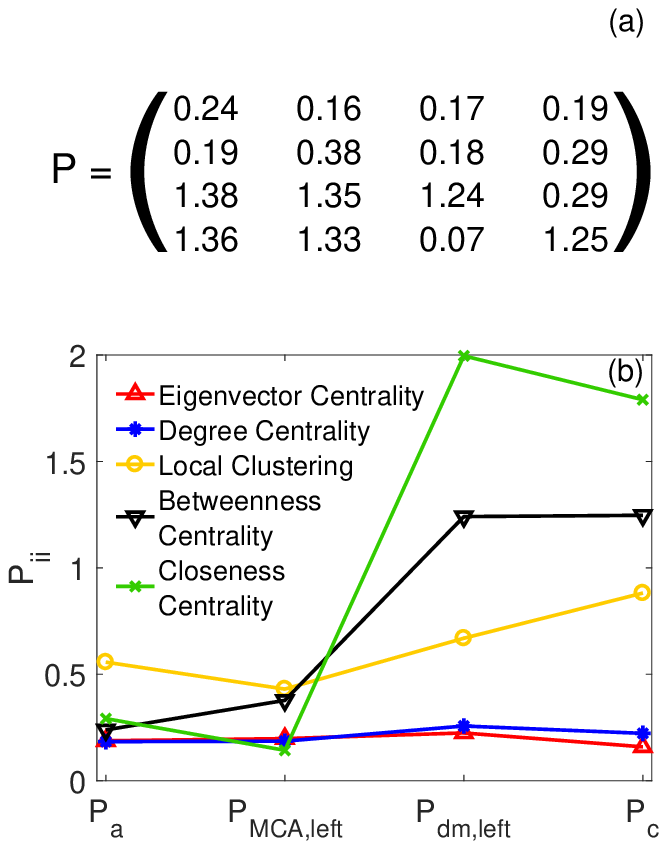}
\caption{(a) Example of PDF difference matrix, $P$, computed over the betweenness centrality, $bc$, for the pressure signal. I row and column: large arteries; II row and column: middle cerebral artery; III row and column: distal middle artery; IV column and row: capillary/venous circulation. (b) Diagonal values of $P$ matrixes for all the metrics considered ($k$, $x$, $bc$, $cc$, $lc$) along the ICA-MCA pathway (pressure signal).}
\end{figure}

\begin{figure*}
\includegraphics[width=1.5\columnwidth]{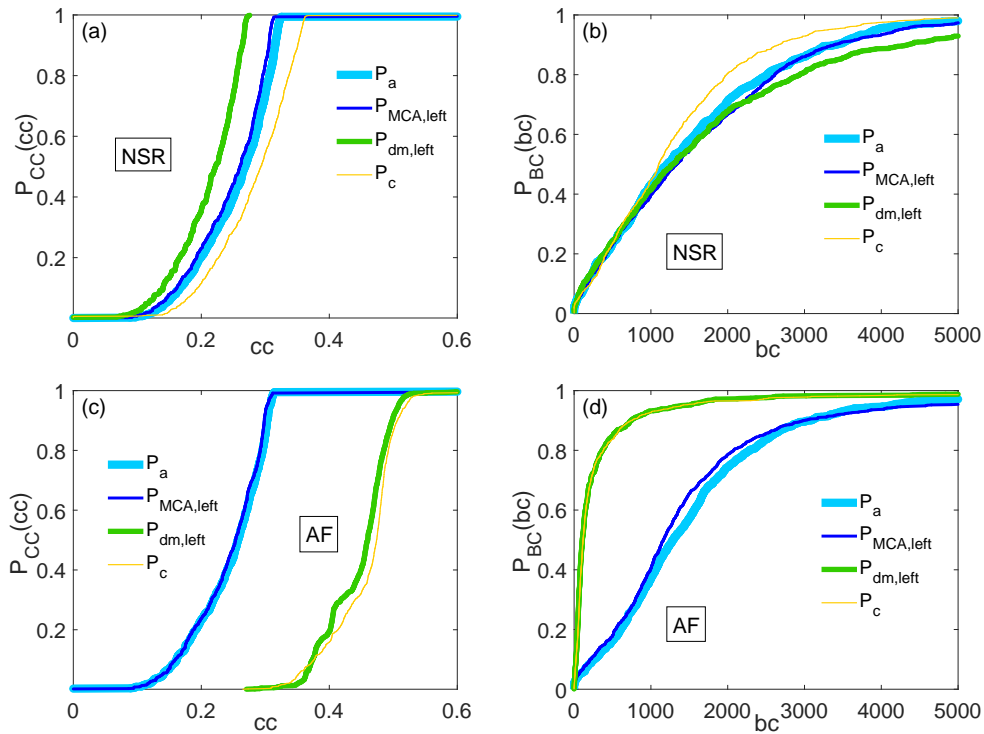}
\caption{Pressure signals. Cumulative density function (CDF) distributions of the closeness centrality (left panels) and betwenness centrality (right panels), during NSR (top panels) and AF (bottom panels) conditions. The four cerebral districts ($P_a$, $P_{MCA,left}$, $P_{dm,left}$, $P_c$) are highlighted with different colors.}
\end{figure*}

A first concise outcome is revealed by the matrix, $P$, defined as
\begin{equation}
P_{ij} = \int_D \left|p_i(x) - p_j(x)\right| dx, \,\,\,\,\,\, i, \,j=1,...,4,
\end{equation}
\noindent where $p_i$ and $p_j$ are two PDFs to be compared, while $D = D_i \cup D_j$ is the union of the domains $D_i$ and $D_j$, where the PDFs are defined. The values of the subscripts $i$ and $j$ vary from 1 to 4 according to the cerebral districts considered (1: large arteries, 2: middle cerebral region, 3: distal region, 4: capillary/venous circulation). $P$, for each metric and hemodynamic signal, accounts for the area of the difference (in module) between pairs of PDF distributions along the different districts in NSR and AF conditions. $P$ has the same dimension [4x4] as the cerebral regions studied. In particular, each element of the diagonal of $P$ represents the area of the difference between two NSR and AF PDFs in the same region (from $P_{11}$ for the large arteries to $P_{44}$ for the capillary/venous circulation). The upper triangular part of $P$ takes into account the difference between two PDFs at different districts in NSR (e.g., $P_{13}$ represents the area of the difference between the PDF at large arteries and the PDF at the distal region during NSR). The lower triangular part of $P$ expresses the difference between two PDFs at different districts in AF (e.g., $P_{24}$ represents the area of the difference between the PDF at the middle cerebral region and the PDF at the capillary/venous region during AF). Since the PDF has unitary area, the area of the difference between the two compared PDFs can vary between 0 (when the distributions coincide) and 2 (when the distributions have completely different domains). An example of $P$ matrix is computed over the betweenness centrality in Fig. 3a for the pressure signal.

\noindent By considering the diagonal values (from top to bottom) of all the $P$ matrices, one can infer how much each metric is significant to capture AF-induced variations (with respect to NSR) along the ICA-MCA pathway, as reported in Fig. 3b. All the metrics studied are displayed for the pressure signal (similar results are found for the flow rate), where on the y-axis lie the $P_{ii}$ values, while on the x-axis the cerebral stations are localized. Fig. 3b shows that the differences between healthy and fibrillated conditions are minimal in large arteries and increase in the distal and capillary circulation. Both the degree and the eigenvector centrality indicators, however, are not optimal metrics in this context, since they do not significantly inherit structural signaling variations induced by AF. The local clustering coefficient, $lc$, together with the closeness, $cc$, and the betweenness, $bc$, centralities are the best metrics to highlight the differences between NSR and AF.

\begin{figure*}
\includegraphics[width=1.5\columnwidth]{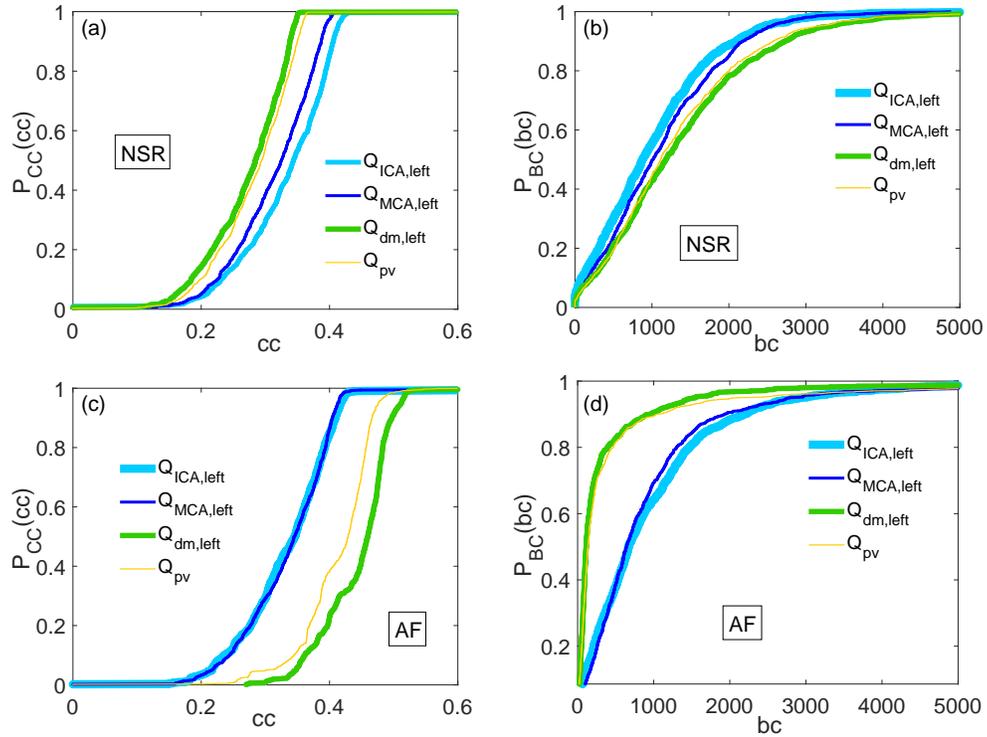}
\caption{Flow rate signals. Cumulative density function (CDF) distributions of the closeness centrality (left panels) and betwenness centrality (right panels), during NSR (top panels) and AF (bottom panels) conditions. The four cerebral districts ($Q_{ICA,left}$, $Q_{MCA,left}$, $Q_{dm,left}$, $Q_{pv}$) are highlighted with different colors.}
\end{figure*}

The analysis of $P$ matrices is exploratory since allows us to discern the most useful metrics from those which here are not so meaningful. To this end, in the following we only focus on $lc$, $bc$ and $cc$ distributions, as well as on the assortative mixing. Moreover, the preliminary examination of $P$ matrices reveals the presence of important variations, in absolute terms, between NSR and AF when going towards the microcirculation. However, evaluating $P$ matrices is not sufficient, as we are not able to observe whether the metrics increase or decrease during AF with respect to NSR along the ICA-MCA pathway. The $bc$ and $cc$ distributions are thus analyzed in a more extensive way through their CDFs. CDFs are shown instead of PDFs because they are less sensitive to oscillations of high-tail values and can be more easily interpreted.

Figure 4 shows the CDFs of the closeness centrality (left) and betweenness centrality (right), starting from the pressures signals. Top and bottom panels refer to NSR and AF conditions, respectively. In each panel the distributions at the four cerebral districts are reported. Figure 5 is organized similarly to Fig. 4, but the results are obtained with the flow rate signals.

\noindent For both pressure and flow rate, during NSR the distributions of $cc$ and $bc$ (top panels of Fig. 4 and 5) assume similar values along the ICA-MCA pathway. During AF, closeness centrality dramatically increases towards the distal and capillary/venous regions (Fig. 4c and 5c), with significantly higher values reached especially for the pressure (Fig. 4c). On the contrary, betweenness centrality in AF tends to meaningfully decrease when entering the microcirculation (Fig. 4d and 5d), for both pressure and flow rate. The two hemodynamic signals (pressure and flow rate) confirm that in normal conditions no relevant variation occurs along the ICA-MCA pathway, while during AF closeness centrality increases and betweenness centrality decreases.

Apart from local clustering and assortativity coefficients, the other metrics ($k$, $x$, $bc$, $cc$) here discussed are measures of a node prominence in a network. However, as observed so far, these measures behave differently and their trends along the cerebral pathway reveal non-trivial behaviors \cite{Valente}. Degree and eigengvector centrality parameters remain almost constant when entering the cerebral regions. As the two metrics represent a similar degree of node centrality, it is reasonable they are quite correlated. On the contrary, closeness and betweenness measures of centrality present opposite trends along the ICA-MCA pathway.

\noindent At the large arteries level in AF conditions, the network has lower $cc$ values and higher $bc$ values with respect to the microcirculation region. Low $cc$ and high $bc$ values mean that the network is topologically elongated and chained. Each node (i.e., beat) is generally linked to the previous and the next nodes (beats), as well as to other far nodes. On average, every node has a low closeness centrality since it has direct links (i.e., low shortest path) only with its neighborhood, while the shortest paths with the rest of the network are in general high. Moreover, each node is equally important with respect to the others for the information transmission. In fact, given two non-directly connected nodes of the network, information has to necessarily pass through the intermediate nodes (beats) between them. This last aspect implies a high $bc$ value for almost all the nodes.

\begin{figure}
\includegraphics[width=0.81\columnwidth]{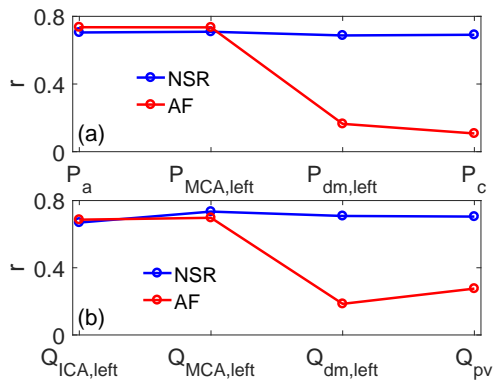}
\caption{Assortativity coefficient, $r$, along the ICA-MCA pathway for  pressure (panel a) and flow rate (panel b) signals. Blue: NSR, red: AF.}
\end{figure}

\noindent In the microcirculation region during AF, the network has higher $cc$ values and lower $bc$ values with respect to the cerebral input. This configuration means that a node is more incline to connect with nodes which are not not its precedent and subsequent nodes. On average, each node needs a limited shortest path to reach all the other nodes of the network. As a consequence, $cc$ value is high. On the contrary, since information has not to cross all the intermediate nodes between pairs of nodes, the betweenness centrality is averagely decreased. Being the shortest paths in general shorter than at large arteries, the stream of information now excludes several nodes which were instead crossed at the carotid entrance. The network is topologically more circular, with non-consecutive distant beats (nodes) which often share a link.

Since during AF degree and eigenvector centrality distributions remain basically constant along the ICA-MCA pathway, this entails that on average each node maintains the same number of links with the other nodes. What makes the difference is the link topology. At the carotid entrance, the links of a node are created with the surrounding nodes as well as with long-range links (beats). Going towards the capillary/venous region, the links connecting consecutive nodes are almost all broken and substituted with long-range links (i.e., links between temporally distant beats).

\begin{figure}
\includegraphics[width=0.81\columnwidth]{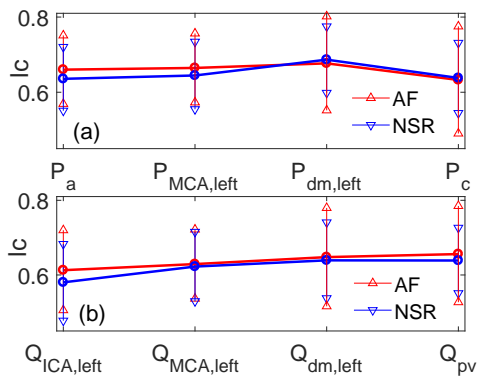}
\caption{Local clustering coefficient, $lc$, along the ICA-MCA pathway for  pressure (panel a) and flow rate (panel b) signals. Mean values $\overline{lc}$, are reported with open circles, while triangles indicate $\overline{lc} \pm \sigma_{lc}$, where $\sigma_{lc}$ is the standard deviation value of the $lc$ distribution. Blue: NSR, red: AF.}
\end{figure}

Evidence of this different link distribution during AF along the cerebral circulation is given by the assortativity coefficient, $r$, along the ICA-MCA pathway (Fig. 6). We recall that $r$ measures the tendency of a network to present link between similar ($r=1$) or dissimilar ($r=-1$) nodes. An assortativity value close to zero means that none of the above trends is evidenced, links emerge with no preference between similar or dissimilar nodes, thus the network is uncorrelated. During NSR, both pressure and flow rate reveal a quite high assortativity (0.65-0.7) which is maintained constant along the ICA-MCA pathway (blue curves, panels a and b). In AF condition, the assortativity has analogous values as in NSR at the carotid entrance, with a significant drop towards the distal and capillary/venous circulation (red curves, panels a and b). In the peripheral regions, links are no more between nodes sharing the same properties (i.e., degree centrality), but spurious long-range links are predominant. The network here resembles the features of a random uncorrelated graph.

\begin{figure*}
\includegraphics[width=1.5\columnwidth]{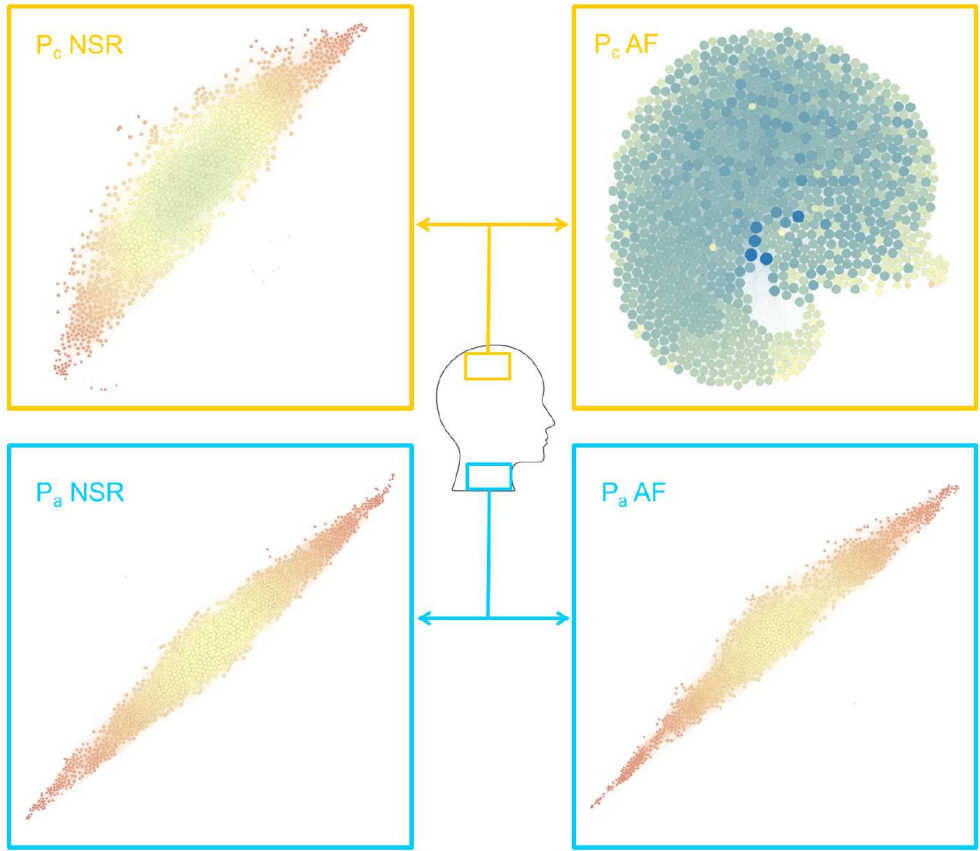}
\caption{Pressure signals. Networks at the internal carotid level (bottom panels, $P_a$) and in the capillary district (top panels, $P_c$), in NSR (left panels) and AF (right panels) conditions. Closeness centrality values are reported through the color (from red, $cc=0$, to blue, $cc=1$) and are expressed as proportional to the size of the nodes.}
\end{figure*}

We conclude the Results Section with a comment on the local clustering coefficient. In Fig. 3b, diagonal elements of the P matrix showed a slight increase of this metric towards the capillary region. In Fig. 7, the mean values, $\overline{lc}$, are reported for both pressure and flow rate signals during NSR and AF, together with the dispersion due to the standard deviation values, $\overline{lc} \pm \sigma_{lc}$, where $\sigma_{lc}$ is the standard deviation value of the $lc$ distribution. It can be noted that, the global local clustering values are quite constant along the ICA-MCA pathway in NSR as well as in AF. However, in the distal and capillary regions, the data dispersion increases (for both NSR and AF), being higher during AF than NSR. Thus, going towards the microcirculation during AF, the neighborhood of each node can be either almost fully connected or poorly connected. This is a further symptom of the increased variability and unpredictability induced by AF on the peripheral hemodynamic signals.

\section{Discussion}

\begin{figure*}
\includegraphics[width=1.5\columnwidth]{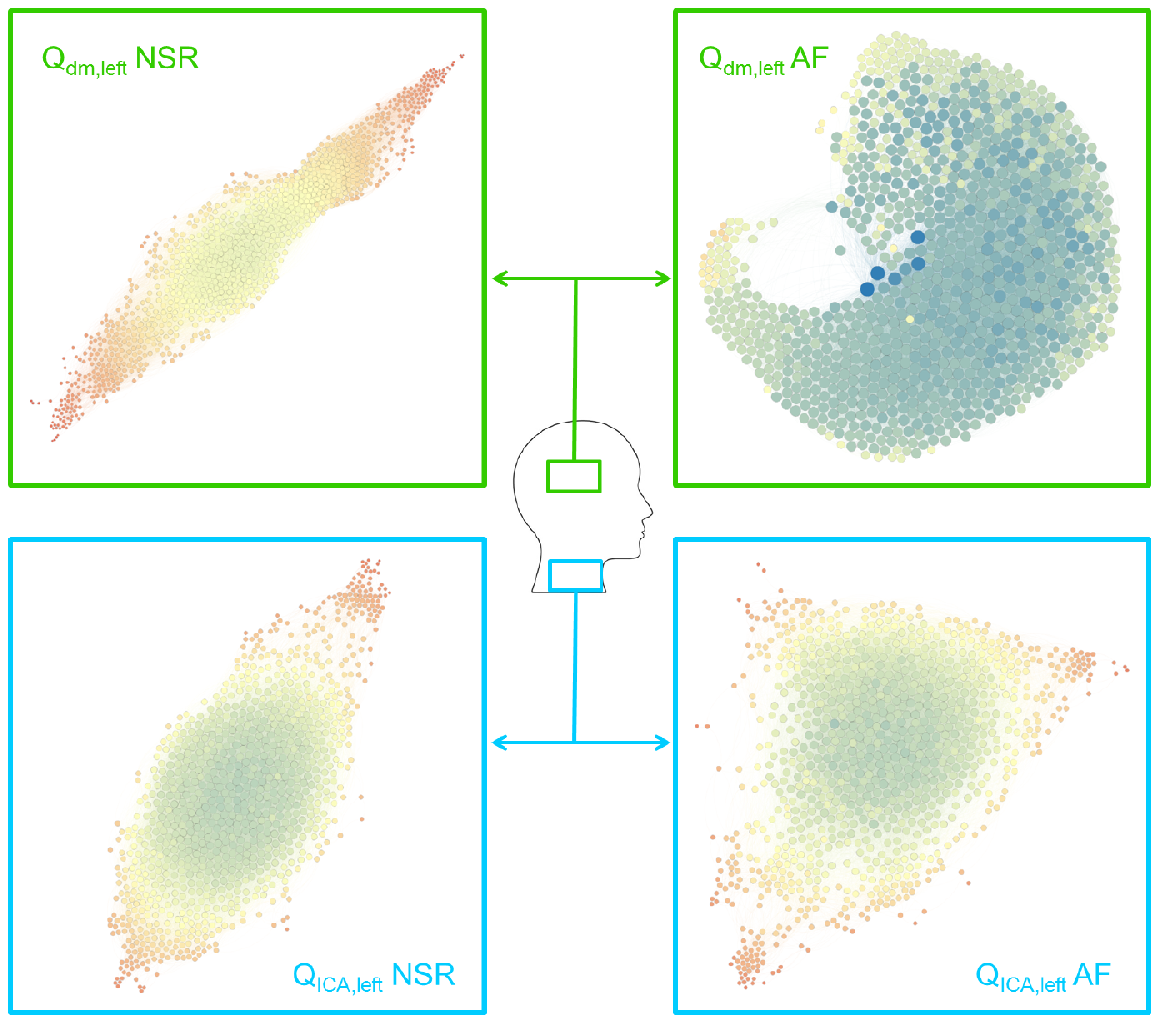}
\caption{Flow rate signals. Networks at the internal carotid level (bottom panels, $Q_{ICA,left}$) and in the distal district (top panels, $Q_{dm,left}$), in NSR (left panels) and AF (right panels) conditions. The size of the nodes is proportional to the closeness centrality value, while color spans from red ($cc=0$) to blue ($cc=1$).}
\end{figure*}

As evidenced in Section III, NSR and AF networks in the cerebral microcirculation present significantly different structures. To better highlight the topological features, we graphically represent the pressure (Fig. 8) and flow rate (Fig. 9) networks at the beginning and towards the end of the ICA-MCA pathway in both NSR and AF conditions. Networks are visualized through the open-source software package \emph{Gephi} \cite{gephi}, with respect to the closeness centrality values. Red color corresponds to low $cc$ values, while blue-colored nodes correspond to high $cc$ values. In addition, the size of a node is proportional to its $cc$ value.

\noindent Let us first consider the networks obtained by the pressure signals, in Fig. 8. At the large arteries level (Fig. 8, bottom panels), the NSR and AF networks present an elongated and almost planar shape, which is the typical feature of networks based on pseudo-periodic series, with similar mean closeness centrality values, $\overline{cc}$ ($\overline{cc}$: 0.257 (NSR), 0.247 (AF)). In the capillary region, the NSR network is slightly less elongated (Fig. 8, top left panel), but the mean closeness centrality value ($\overline{cc}= 0.282$) is not far from the corresponding value at the entrance ($\overline{cc}=0.257$). During AF the scenario is instead completely altered (Fig. 8, top right panel). The network assumes a markedly circular and three-dimensional shape, which is usually encountered in random networks, with an average closeness centrality value ($\overline{cc}=0.458$) which almost doubles the corresponding one at the cerebral entrance ($\overline{cc}=0.247$).

The networks obtained from flow rate signals have similar average closeness centrality values at the large arteries level (Fig. 9, bottom panels), for both NSR ($\overline{cc}= 0.333$) and AF ($\overline{cc}= 0.337$). The NSR network (Fig. 9, bottom left panel) has an elongated shape, although less pronounced than the corresponding pressure NSR network (Fig. 8, bottom left panel). During AF, the flow rate network assumes a triangular shape, preserving its bi-dimensional form (Fig. 9, bottom right panel). In the distal region, the NSR network maintains its mainly-planar shape (Fig. 9, top left panel) and accentuates its elongation with respect to the carotid entrance. The average closeness centrality value ($\overline{cc}=0.284$) does not change significantly with respect to the $Q_{ICA,left}$ network ($\overline{cc}=0.333$). On the contrary, in AF the network becomes circular and fully three-dimensional (Fig. 9, top right panel), with an increased closeness centrality mean value ($\overline{cc}=0.412$). In this configuration, the network features resemble those of random networks.

The network-based approach is able to fully catch the structural differences between NSR and AF hemodynamic signals, which, for some metrics, are striking. The proposed scenario evidences a dramatic signal variation towards the distal/capillary cerebral regions during AF, which has no counterpart in NSR conditions. All the most significant network parameters ($bc$, $cc$, $lc$, $r$) agree in locating this alteration. During AF, the input signals ($P_a$ and $Q_{ICA,left}$) exhibit pseudo-periodic features, which are almost completely lost towards the microcirculation, where instead $P_c$ and $Q_{dm,left}$ signals reveal random-like characteristics.

\noindent The hemodynamic consequences of this substantial alteration can be of high biomedical impact. All the physiological phenomena at microcerebral level ruled by periodicity - such as regular perfusion, mean pressure per beat, average nutrient supply at cellular level - are not guaranteed and can be strongly compromised, since the AF hemodynamic signals assume irregular behaviour and random-like features. The cardiovascular implications of the highlighted alteration surely deserve to be further quantified through clinical evidences, although invasive and accurate measurements are still difficult to be accomplished, also because of the signal complexity induced by the heart rate variability. Awaiting necessary \textit{in vivo} validation, the network-based hints here emerged can plausibly explain the hemodynamic mechanisms leading to cognitive impairment in presence of persistent AF.


\section{Conclusions}

The network analysis performed over cerebral hemodynamic signals highlighted that the degree centrality, which is usually the most intuitive and firstly analyzed metric, is here not much informative. Other local (i.e., eigenvector centrality) and mesoscopic (i.e., local clustering coefficient) measures are not crucial in discerning NSR and AF hemodynamic features. A deeper examination of the adjacency matrix was necessary, especially in terms of global metrics. In particular, the markers of betweenness and closeness centrality as well as the assortativity coefficient turned out to be meaningful. From the combined analysis of the network metrics through their probability distributions, during AF it emerges that towards the peripheral cerebral circulation the closeness centrality increases, while the betweeness centrality is reduced. The AF pressure and flow rate networks change from an elongated shape (which is characteristic of pseudo-periodic series) in the large artery region to a circular-like shape (which is a feature of random series) in the capillary-venous districts.

The complex network analysis evidences in a synthetic and innovative way how hemodynamic signals in the cerebral microcirculation are deeply altered by AF. This result, on one hand, confirms that the complex network theory can be successfully extended to explore other pathological biomedical signals in complex geometries, such as stenotic flows across aortic valve and flow dynamics in brain aneurysms. On the other hand, the present findings further strengthen the always more evident link between AF hemodynamic and cognitive decline, through a powerful approach which is complementary to the classical statistical tools.

\appendix
\section{Sensitivity Analysis}

\renewcommand{\thefigure}{A\arabic{figure}}
\setcounter{figure}{0}
\begin{figure}
\includegraphics[width=0.81\columnwidth]{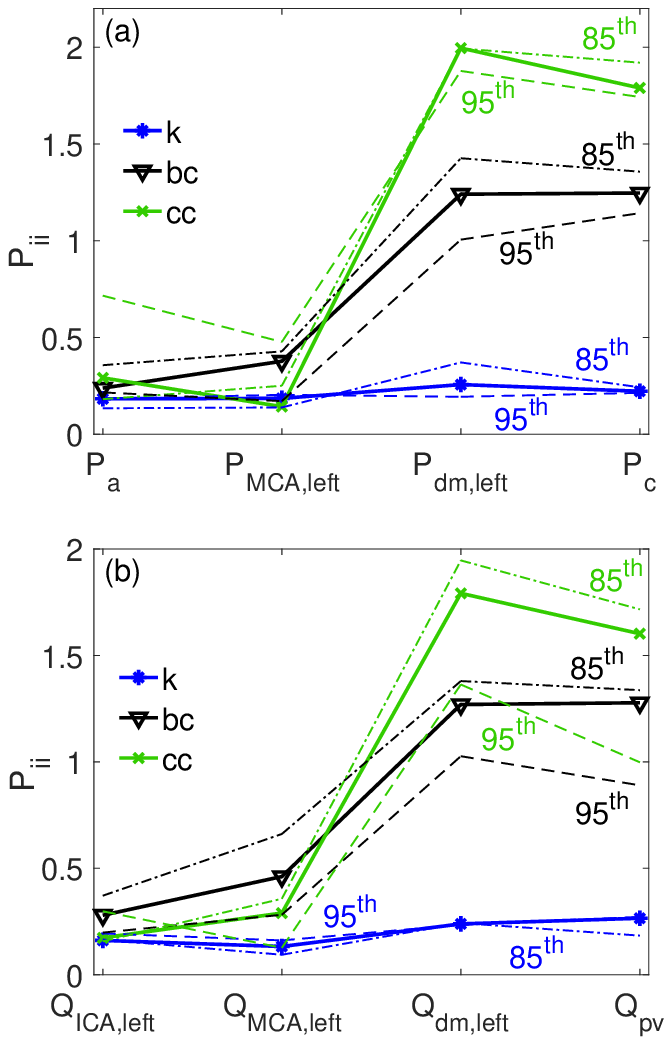}
\caption{Diagonal values of $P$ matrixes for the degree, closeness and betweenness centrality measures ($k$, $cc$, $bc$) along the ICA-MCA pathway considering three percentile values: $85^{th}$ (dashed-dotted line), $90^{th}$ (thick line with symbols), $95^{th}$ (dashed line). (a) pressure, (b) flow rate.}
\end{figure}

A sensitivity analysis on the threshold value $\tau$ is here performed, recalling that $\tau$ is the only arbitrary parameter involved for building the network. Beside the $90^{th}$ percentile of the $R$ matrix, other two values have been chosen to set the threshold $\tau$, namely the $85^{th}$ and $95^{th}$ percentiles. In Fig. A1, the diagonal of $P$ matrix is reported for the degree centrality, closeness centrality and betweenness centrality for the three percentile values, for both pressure (top panel) and flow rate (bottom panel) signals. It can be noted that, apart from the specific values, the trend of the $P_{ii}$ values remain unaltered by changing the threshold $\tau$. In fact, for all the percentile values considered, the degree centrality, $k$, remains almost constant towards the peripheral circulation, while betweenness and closeness centrality values experience significative variations with respect to large arteries. The proposed percentile ($90^{th}$) is therefore a good choice to catch the network behavior which, despite the specific values assumed at each cerebral district, turns out to be substantially insensitive to the threshold adopted.

\nocite{*}

\bibliography{scarsoglio}

\providecommand{\noopsort}[1]{}\providecommand{\singleletter}[1]{#1}%
\begin{thebibliography}{58}%
\makeatletter
\providecommand \@ifxundefined [1]{%
 \@ifx{#1\undefined}
}%
\providecommand \@ifnum [1]{%
 \ifnum #1\expandafter \@firstoftwo
 \else \expandafter \@secondoftwo
 \fi
}%
\providecommand \@ifx [1]{%
 \ifx #1\expandafter \@firstoftwo
 \else \expandafter \@secondoftwo
 \fi
}%
\providecommand \natexlab [1]{#1}%
\providecommand \enquote  [1]{``#1''}%
\providecommand \bibnamefont  [1]{#1}%
\providecommand \bibfnamefont [1]{#1}%
\providecommand \citenamefont [1]{#1}%
\providecommand \href@noop [0]{\@secondoftwo}%
\providecommand \href [0]{\begingroup \@sanitize@url \@href}%
\providecommand \@href[1]{\@@startlink{#1}\@@href}%
\providecommand \@@href[1]{\endgroup#1\@@endlink}%
\providecommand \@sanitize@url [0]{\catcode `\\12\catcode `\$12\catcode
  `\&12\catcode `\#12\catcode `\^12\catcode `\_12\catcode `\%12\relax}%
\providecommand \@@startlink[1]{}%
\providecommand \@@endlink[0]{}%
\providecommand \url  [0]{\begingroup\@sanitize@url \@url }%
\providecommand \@url [1]{\endgroup\@href {#1}{\urlprefix }}%
\providecommand \urlprefix  [0]{URL }%
\providecommand \Eprint [0]{\href }%
\providecommand \doibase [0]{http://dx.doi.org/}%
\providecommand \selectlanguage [0]{\@gobble}%
\providecommand \bibinfo  [0]{\@secondoftwo}%
\providecommand \bibfield  [0]{\@secondoftwo}%
\providecommand \translation [1]{[#1]}%
\providecommand \BibitemOpen [0]{}%
\providecommand \bibitemStop [0]{}%
\providecommand \bibitemNoStop [0]{.\EOS\space}%
\providecommand \EOS [0]{\spacefactor3000\relax}%
\providecommand \BibitemShut  [1]{\csname bibitem#1\endcsname}%
\let\auto@bib@innerbib\@empty
\bibitem [{\citenamefont {Etchings}\ and\ \citenamefont
  {Buetow}(2017)}]{Etchings}%
  \BibitemOpen
  \bibfield  {author} {\bibinfo {author} {\bibfnamefont {J.~A.}\ \bibnamefont
  {Etchings}}\ and\ \bibinfo {author} {\bibfnamefont {K.}~\bibnamefont
  {Buetow}},\ }\href@noop {} {\emph {\bibinfo {title} {Strategies in Biomedical
  Data Science: Driving Force for Innovation}}}\ (\bibinfo  {publisher}
  {Wiley},\ \bibinfo {year} {2017})\BibitemShut {NoStop}%
\bibitem [{\citenamefont {Albert}\ and\ \citenamefont
  {Barab{\'{a}}si}(2002)}]{albert_barabasi_2002}%
  \BibitemOpen
  \bibfield  {author} {\bibinfo {author} {\bibfnamefont {R.}~\bibnamefont
  {Albert}}\ and\ \bibinfo {author} {\bibfnamefont {A.~L.}\ \bibnamefont
  {Barab{\'{a}}si}},\ }\bibfield  {title} {\enquote {\bibinfo {title}
  {Statistical mechanics of complex networks},}\ }\href@noop {} {\bibfield
  {journal} {\bibinfo  {journal} {Review of Modern Physics}\ }\textbf {\bibinfo
  {volume} {74}},\ \bibinfo {pages} {47--97} (\bibinfo {year}
  {2002})}\BibitemShut {NoStop}%
\bibitem [{\citenamefont {Boccaletti}\ \emph {et~al.}(2006)\citenamefont
  {Boccaletti}, \citenamefont {Latora}, \citenamefont {Moreno}, \citenamefont
  {Chavez},\ and\ \citenamefont {Hwang}}]{Boccaletti_et_al_2006}%
  \BibitemOpen
  \bibfield  {author} {\bibinfo {author} {\bibfnamefont {S.}~\bibnamefont
  {Boccaletti}}, \bibinfo {author} {\bibfnamefont {V.}~\bibnamefont {Latora}},
  \bibinfo {author} {\bibfnamefont {Y.}~\bibnamefont {Moreno}}, \bibinfo
  {author} {\bibfnamefont {M.}~\bibnamefont {Chavez}}, \ and\ \bibinfo {author}
  {\bibfnamefont {D.~U.}\ \bibnamefont {Hwang}},\ }\bibfield  {title} {\enquote
  {\bibinfo {title} {Complex networks: Structure and dynamics},}\ }\href@noop
  {} {\bibfield  {journal} {\bibinfo  {journal} {Physics Reports}\ }\textbf
  {\bibinfo {volume} {424}},\ \bibinfo {pages} {175--308} (\bibinfo {year}
  {2006})}\BibitemShut {NoStop}%
\bibitem [{\citenamefont {Watts}\ and\ \citenamefont
  {Strogatz}(1998)}]{WS1998}%
  \BibitemOpen
  \bibfield  {author} {\bibinfo {author} {\bibfnamefont {D.~J.}\ \bibnamefont
  {Watts}}\ and\ \bibinfo {author} {\bibfnamefont {S.~H.}\ \bibnamefont
  {Strogatz}},\ }\bibfield  {title} {\enquote {\bibinfo {title} {Collective
  dynamics of 'small-world' networks},}\ }\href@noop {} {\bibfield  {journal}
  {\bibinfo  {journal} {Nature}\ }\textbf {\bibinfo {volume} {393}},\ \bibinfo
  {pages} {440–--442} (\bibinfo {year} {1998})}\BibitemShut {NoStop}%
\bibitem [{\citenamefont {Costa}\ \emph {et~al.}(2011)\citenamefont {Costa},
  \citenamefont {Oliveira}, \citenamefont {Travieso}, \citenamefont
  {Rodrigues}, \citenamefont {Boas}, \citenamefont {Antiqueira}, \citenamefont
  {Viana},\ and\ \citenamefont {Rocha}}]{costa_et_al_2011}%
  \BibitemOpen
  \bibfield  {author} {\bibinfo {author} {\bibfnamefont {L.~D.~F.}\
  \bibnamefont {Costa}}, \bibinfo {author} {\bibfnamefont {O.~N.}\ \bibnamefont
  {Oliveira}}, \bibinfo {author} {\bibfnamefont {G.}~\bibnamefont {Travieso}},
  \bibinfo {author} {\bibfnamefont {F.~A.}\ \bibnamefont {Rodrigues}}, \bibinfo
  {author} {\bibfnamefont {P.~R.~V.}\ \bibnamefont {Boas}}, \bibinfo {author}
  {\bibfnamefont {L.}~\bibnamefont {Antiqueira}}, \bibinfo {author}
  {\bibfnamefont {M.~P.}\ \bibnamefont {Viana}}, \ and\ \bibinfo {author}
  {\bibfnamefont {L.~E.~C.}\ \bibnamefont {Rocha}},\ }\bibfield  {title}
  {\enquote {\bibinfo {title} {Analyzing and modeling real-world phenomena with
  complex networks: a survey of applications},}\ }\href@noop {} {\bibfield
  {journal} {\bibinfo  {journal} {Advances in Physics}\ }\textbf {\bibinfo
  {volume} {60}},\ \bibinfo {pages} {329--412} (\bibinfo {year}
  {2011})}\BibitemShut {NoStop}%
\bibitem [{\citenamefont {Havlin}\ \emph {et~al.}(2012)\citenamefont {Havlin},
  \citenamefont {Kenett}, \citenamefont {Ben-Jacob}, \citenamefont {Bunde},
  \citenamefont {Cohen}, \citenamefont {Hermann}, \citenamefont {Kantelhardt},
  \citenamefont {Kertész}, \citenamefont {Kirkpatrick}, \citenamefont
  {Kurths}, \citenamefont {Portugali},\ and\ \citenamefont {Solomon}}]{Havlin}%
  \BibitemOpen
  \bibfield  {author} {\bibinfo {author} {\bibfnamefont {S.}~\bibnamefont
  {Havlin}}, \bibinfo {author} {\bibfnamefont {D.~Y.}\ \bibnamefont {Kenett}},
  \bibinfo {author} {\bibfnamefont {E.}~\bibnamefont {Ben-Jacob}}, \bibinfo
  {author} {\bibfnamefont {A.}~\bibnamefont {Bunde}}, \bibinfo {author}
  {\bibfnamefont {R.}~\bibnamefont {Cohen}}, \bibinfo {author} {\bibfnamefont
  {H.}~\bibnamefont {Hermann}}, \bibinfo {author} {\bibfnamefont {J.~W.}\
  \bibnamefont {Kantelhardt}}, \bibinfo {author} {\bibfnamefont
  {J.}~\bibnamefont {Kertész}}, \bibinfo {author} {\bibfnamefont
  {S.}~\bibnamefont {Kirkpatrick}}, \bibinfo {author} {\bibfnamefont
  {J.}~\bibnamefont {Kurths}}, \bibinfo {author} {\bibfnamefont
  {J.}~\bibnamefont {Portugali}}, \ and\ \bibinfo {author} {\bibfnamefont
  {S.}~\bibnamefont {Solomon}},\ }\bibfield  {title} {\enquote {\bibinfo
  {title} {Challenges in network science: Applications to infrastructures,
  climate, social systems and economics},}\ }\href@noop {} {\bibfield
  {journal} {\bibinfo  {journal} {European Physical Journal Special Topics}\
  }\textbf {\bibinfo {volume} {214}},\ \bibinfo {pages} {273–--293} (\bibinfo
  {year} {2012})}\BibitemShut {NoStop}%
\bibitem [{\citenamefont {Fang}, \citenamefont {Sivakumar},\ and\ \citenamefont
  {Woldemeskel}(2017)}]{Sivakumar}%
  \BibitemOpen
  \bibfield  {author} {\bibinfo {author} {\bibfnamefont {K.}~\bibnamefont
  {Fang}}, \bibinfo {author} {\bibfnamefont {B.}~\bibnamefont {Sivakumar}}, \
  and\ \bibinfo {author} {\bibfnamefont {F.~M.}\ \bibnamefont {Woldemeskel}},\
  }\bibfield  {title} {\enquote {\bibinfo {title} {Complex networks, community
  structure, and catchment classification in a large-scale river basin},}\
  }\href@noop {} {\bibfield  {journal} {\bibinfo  {journal} {Journal of
  Hydrology}\ }\textbf {\bibinfo {volume} {545}},\ \bibinfo {pages} {478--493}
  (\bibinfo {year} {2017})}\BibitemShut {NoStop}%
\bibitem [{\citenamefont {Donges}\ \emph {et~al.}(2009)\citenamefont {Donges},
  \citenamefont {Zou}, \citenamefont {Marwan},\ and\ \citenamefont
  {Kurths}}]{Donges_et_al_2009}%
  \BibitemOpen
  \bibfield  {author} {\bibinfo {author} {\bibfnamefont {J.~F.}\ \bibnamefont
  {Donges}}, \bibinfo {author} {\bibfnamefont {Y.}~\bibnamefont {Zou}},
  \bibinfo {author} {\bibfnamefont {N.}~\bibnamefont {Marwan}}, \ and\ \bibinfo
  {author} {\bibfnamefont {J.}~\bibnamefont {Kurths}},\ }\bibfield  {title}
  {\enquote {\bibinfo {title} {Complex networks in climate dynamics},}\
  }\href@noop {} {\bibfield  {journal} {\bibinfo  {journal} {European Physical
  Journal Special Topics}\ }\textbf {\bibinfo {volume} {174}},\ \bibinfo
  {pages} {157--179} (\bibinfo {year} {2009})}\BibitemShut {NoStop}%
\bibitem [{\citenamefont {Scarsoglio}, \citenamefont {Laio},\ and\
  \citenamefont {Ridolfi}(2013)}]{scarsoglio2013}%
  \BibitemOpen
  \bibfield  {author} {\bibinfo {author} {\bibfnamefont {S.}~\bibnamefont
  {Scarsoglio}}, \bibinfo {author} {\bibfnamefont {F.}~\bibnamefont {Laio}}, \
  and\ \bibinfo {author} {\bibfnamefont {L.}~\bibnamefont {Ridolfi}},\
  }\bibfield  {title} {\enquote {\bibinfo {title} {Climate dynamics: a
  network-based approach for the analysis of global precipitation},}\
  }\href@noop {} {\bibfield  {journal} {\bibinfo  {journal} {PLoS ONE}\
  }\textbf {\bibinfo {volume} {8}},\ \bibinfo {pages} {e71129} (\bibinfo {year}
  {2013})}\BibitemShut {NoStop}%
\bibitem [{\citenamefont {Tsonis}\ and\ \citenamefont
  {Swanson}(2008)}]{Tsonis_Swanson_2008}%
  \BibitemOpen
  \bibfield  {author} {\bibinfo {author} {\bibfnamefont {A.~A.}\ \bibnamefont
  {Tsonis}}\ and\ \bibinfo {author} {\bibfnamefont {K.~L.}\ \bibnamefont
  {Swanson}},\ }\bibfield  {title} {\enquote {\bibinfo {title} {Topology and
  predictability of el ni$\tilde{\textmd{n}}$o and la ni$\tilde{\textmd{n}}$a
  networks},}\ }\href@noop {} {\bibfield  {journal} {\bibinfo  {journal}
  {Physical Review Letters}\ }\textbf {\bibinfo {volume} {100}},\ \bibinfo
  {pages} {228502} (\bibinfo {year} {2008})}\BibitemShut {NoStop}%
\bibitem [{\citenamefont {Taira}, \citenamefont {Nair},\ and\ \citenamefont
  {Brunton}(2016)}]{Taira}%
  \BibitemOpen
  \bibfield  {author} {\bibinfo {author} {\bibfnamefont {K.}~\bibnamefont
  {Taira}}, \bibinfo {author} {\bibfnamefont {A.~G.}\ \bibnamefont {Nair}}, \
  and\ \bibinfo {author} {\bibfnamefont {S.~L.}\ \bibnamefont {Brunton}},\
  }\bibfield  {title} {\enquote {\bibinfo {title} {Network structure of
  two-dimensional decaying isotropic turbulence},}\ }\href@noop {} {\bibfield
  {journal} {\bibinfo  {journal} {Journal of Fluid Mechanics}\ }\textbf
  {\bibinfo {volume} {795}},\ \bibinfo {pages} {795R21} (\bibinfo {year}
  {2016})}\BibitemShut {NoStop}%
\bibitem [{\citenamefont {Scarsoglio}, \citenamefont {Iacobello},\ and\
  \citenamefont {Ridolfi}(2016)}]{Scarsoglio_IJBC_2016}%
  \BibitemOpen
  \bibfield  {author} {\bibinfo {author} {\bibfnamefont {S.}~\bibnamefont
  {Scarsoglio}}, \bibinfo {author} {\bibfnamefont {G.}~\bibnamefont
  {Iacobello}}, \ and\ \bibinfo {author} {\bibfnamefont {L.}~\bibnamefont
  {Ridolfi}},\ }\bibfield  {title} {\enquote {\bibinfo {title} {Complex
  networks unveiling spatial patterns in turbulence},}\ }\href@noop {}
  {\bibfield  {journal} {\bibinfo  {journal} {International Journal of
  Bifurcation and Chaos}\ }\textbf {\bibinfo {volume} {26}},\ \bibinfo {pages}
  {1650223} (\bibinfo {year} {2016})}\BibitemShut {NoStop}%
\bibitem [{\citenamefont {Tupikina}\ \emph {et~al.}(2016)\citenamefont
  {Tupikina}, \citenamefont {Molkenthin}, \citenamefont
  {L$\acute{\textmd{o}}$pez}, \citenamefont
  {E.~Hern$\acute{\textmd{a}}$ndez-Garc$\acute{\textmd{i}}$a},\ and\
  \citenamefont {Kurths}}]{Tupikina}%
  \BibitemOpen
  \bibfield  {author} {\bibinfo {author} {\bibfnamefont {L.}~\bibnamefont
  {Tupikina}}, \bibinfo {author} {\bibfnamefont {N.}~\bibnamefont
  {Molkenthin}}, \bibinfo {author} {\bibfnamefont {C.}~\bibnamefont
  {L$\acute{\textmd{o}}$pez}}, \bibinfo {author} {\bibfnamefont {N.~M.}\
  \bibnamefont {E.~Hern$\acute{\textmd{a}}$ndez-Garc$\acute{\textmd{i}}$a}}, \
  and\ \bibinfo {author} {\bibfnamefont {J.}~\bibnamefont {Kurths}},\
  }\bibfield  {title} {\enquote {\bibinfo {title} {Correlation networks from
  flows. the case of forced and time-dependent advection-diffusion dynamics},}\
  }\href@noop {} {\bibfield  {journal} {\bibinfo  {journal} {PLoS ONE}\
  }\textbf {\bibinfo {volume} {11}},\ \bibinfo {pages} {e0153703} (\bibinfo
  {year} {2016})}\BibitemShut {NoStop}%
\bibitem [{\citenamefont {Reijneveld}\ \emph {et~al.}(2007)\citenamefont
  {Reijneveld}, \citenamefont {Ponten}, \citenamefont {Berendse},\ and\
  \citenamefont {Stam}}]{Reijneveld}%
  \BibitemOpen
  \bibfield  {author} {\bibinfo {author} {\bibfnamefont {J.~C.}\ \bibnamefont
  {Reijneveld}}, \bibinfo {author} {\bibfnamefont {S.~C.}\ \bibnamefont
  {Ponten}}, \bibinfo {author} {\bibfnamefont {H.~W.}\ \bibnamefont
  {Berendse}}, \ and\ \bibinfo {author} {\bibfnamefont {C.~J.}\ \bibnamefont
  {Stam}},\ }\bibfield  {title} {\enquote {\bibinfo {title} {The application of
  graph theoretical analysis to complex networks in the brain},}\ }\href@noop
  {} {\bibfield  {journal} {\bibinfo  {journal} {Clinical Neurophysiology}\
  }\textbf {\bibinfo {volume} {118}},\ \bibinfo {pages} {2317–--2331}
  (\bibinfo {year} {2007})}\BibitemShut {NoStop}%
\bibitem [{\citenamefont {Lusis}\ and\ \citenamefont {Weiss}(2010)}]{Lusis}%
  \BibitemOpen
  \bibfield  {author} {\bibinfo {author} {\bibfnamefont {A.~J.}\ \bibnamefont
  {Lusis}}\ and\ \bibinfo {author} {\bibfnamefont {J.~N.}\ \bibnamefont
  {Weiss}},\ }\bibfield  {title} {\enquote {\bibinfo {title} {Cardiovascular
  networks: Systems-based approaches to cardiovascular disease},}\ }\href@noop
  {} {\bibfield  {journal} {\bibinfo  {journal} {Circulation}\ }\textbf
  {\bibinfo {volume} {121}},\ \bibinfo {pages} {157--170} (\bibinfo {year}
  {2010})}\BibitemShut {NoStop}%
\bibitem [{\citenamefont {$\acute{\textmd{A}}$vila}\ \emph
  {et~al.}(2013)\citenamefont {$\acute{\textmd{A}}$vila}, \citenamefont
  {Gapelyuk}, \citenamefont {Marwan}, \citenamefont {Walther}, \citenamefont
  {Stepan}, \citenamefont {Kurths},\ and\ \citenamefont {Wessel}}]{Avila}%
  \BibitemOpen
  \bibfield  {author} {\bibinfo {author} {\bibfnamefont {G.~M.~R.}\
  \bibnamefont {$\acute{\textmd{A}}$vila}}, \bibinfo {author} {\bibfnamefont
  {A.}~\bibnamefont {Gapelyuk}}, \bibinfo {author} {\bibfnamefont
  {N.}~\bibnamefont {Marwan}}, \bibinfo {author} {\bibfnamefont
  {T.}~\bibnamefont {Walther}}, \bibinfo {author} {\bibfnamefont
  {H.}~\bibnamefont {Stepan}}, \bibinfo {author} {\bibfnamefont
  {J.}~\bibnamefont {Kurths}}, \ and\ \bibinfo {author} {\bibfnamefont
  {N.}~\bibnamefont {Wessel}},\ }\bibfield  {title} {\enquote {\bibinfo {title}
  {Classification of cardiovascular time series based on different coupling
  structures using recurrence networks analysis},}\ }\href@noop {} {\bibfield
  {journal} {\bibinfo  {journal} {Philosophical Transactions of the Royal
  Society A}\ }\textbf {\bibinfo {volume} {371}},\ \bibinfo {pages} {20110623}
  (\bibinfo {year} {2013})}\BibitemShut {NoStop}%
\bibitem [{\citenamefont {Bullmore}\ and\ \citenamefont
  {Sporns}(2009)}]{Bullmore}%
  \BibitemOpen
  \bibfield  {author} {\bibinfo {author} {\bibfnamefont {E.}~\bibnamefont
  {Bullmore}}\ and\ \bibinfo {author} {\bibfnamefont {O.}~\bibnamefont
  {Sporns}},\ }\bibfield  {title} {\enquote {\bibinfo {title} {Complex brain
  networks: graph theoretical analysis of structural and functional systems},}\
  }\href@noop {} {\bibfield  {journal} {\bibinfo  {journal} {Nature Reviews
  Neuroscience}\ }\textbf {\bibinfo {volume} {10}},\ \bibinfo {pages}
  {186--198} (\bibinfo {year} {2009})}\BibitemShut {NoStop}%
\bibitem [{\citenamefont {Donner}\ \emph {et~al.}(2011)\citenamefont {Donner},
  \citenamefont {Small}, \citenamefont {Donges}, \citenamefont {Marwan},
  \citenamefont {Zou}, \citenamefont {Xiang},\ and\ \citenamefont
  {Kurths}}]{Donner}%
  \BibitemOpen
  \bibfield  {author} {\bibinfo {author} {\bibfnamefont {R.~V.}\ \bibnamefont
  {Donner}}, \bibinfo {author} {\bibfnamefont {M.}~\bibnamefont {Small}},
  \bibinfo {author} {\bibfnamefont {J.~F.}\ \bibnamefont {Donges}}, \bibinfo
  {author} {\bibfnamefont {N.}~\bibnamefont {Marwan}}, \bibinfo {author}
  {\bibfnamefont {Y.}~\bibnamefont {Zou}}, \bibinfo {author} {\bibfnamefont
  {R.}~\bibnamefont {Xiang}}, \ and\ \bibinfo {author} {\bibfnamefont
  {J.}~\bibnamefont {Kurths}},\ }\bibfield  {title} {\enquote {\bibinfo {title}
  {Recurrence-based time series analysis by means of complex network
  methods},}\ }\href@noop {} {\bibfield  {journal} {\bibinfo  {journal}
  {International Journal of Bifurcattion and Chaos}\ }\textbf {\bibinfo
  {volume} {21}},\ \bibinfo {pages} {1019--1046} (\bibinfo {year}
  {2011})}\BibitemShut {NoStop}%
\bibitem [{\citenamefont {Lacasa}\ \emph {et~al.}(2008)\citenamefont {Lacasa},
  \citenamefont {Luque}, \citenamefont {Ballesteros}, \citenamefont {Luque},\
  and\ \citenamefont {Nuno}}]{Lacasa}%
  \BibitemOpen
  \bibfield  {author} {\bibinfo {author} {\bibfnamefont {L.}~\bibnamefont
  {Lacasa}}, \bibinfo {author} {\bibfnamefont {B.}~\bibnamefont {Luque}},
  \bibinfo {author} {\bibfnamefont {F.}~\bibnamefont {Ballesteros}}, \bibinfo
  {author} {\bibfnamefont {J.}~\bibnamefont {Luque}}, \ and\ \bibinfo {author}
  {\bibfnamefont {J.~C.}\ \bibnamefont {Nuno}},\ }\bibfield  {title} {\enquote
  {\bibinfo {title} {From time series to complex networks: The visibility
  graph},}\ }\href@noop {} {\bibfield  {journal} {\bibinfo  {journal}
  {Proceedings of the National Academy of Sciences USA}\ }\textbf {\bibinfo
  {volume} {105}},\ \bibinfo {pages} {4972--4975} (\bibinfo {year}
  {2008})}\BibitemShut {NoStop}%
\bibitem [{\citenamefont {Zhang}\ and\ \citenamefont
  {Small}(2006)}]{Zhang_2006}%
  \BibitemOpen
  \bibfield  {author} {\bibinfo {author} {\bibfnamefont {J.}~\bibnamefont
  {Zhang}}\ and\ \bibinfo {author} {\bibfnamefont {M.}~\bibnamefont {Small}},\
  }\bibfield  {title} {\enquote {\bibinfo {title} {Complex network from
  pseudoperiodic time series: Topology versus dynamics},}\ }\href@noop {}
  {\bibfield  {journal} {\bibinfo  {journal} {Physical Review Letters}\
  }\textbf {\bibinfo {volume} {96}},\ \bibinfo {pages} {238701} (\bibinfo
  {year} {2006})}\BibitemShut {NoStop}%
\bibitem [{\citenamefont {Zhang}\ \emph {et~al.}(2008)\citenamefont {Zhang},
  \citenamefont {Sun}, \citenamefont {Luo}, \citenamefont {Zhang},
  \citenamefont {Nakamura},\ and\ \citenamefont {Small}}]{Zhang_et_al_2008}%
  \BibitemOpen
  \bibfield  {author} {\bibinfo {author} {\bibfnamefont {J.}~\bibnamefont
  {Zhang}}, \bibinfo {author} {\bibfnamefont {J.}~\bibnamefont {Sun}}, \bibinfo
  {author} {\bibfnamefont {X.}~\bibnamefont {Luo}}, \bibinfo {author}
  {\bibfnamefont {K.}~\bibnamefont {Zhang}}, \bibinfo {author} {\bibfnamefont
  {T.}~\bibnamefont {Nakamura}}, \ and\ \bibinfo {author} {\bibfnamefont
  {M.}~\bibnamefont {Small}},\ }\bibfield  {title} {\enquote {\bibinfo {title}
  {Characterizing pseudoperiodic time series through the complex network
  approach},}\ }\href@noop {} {\bibfield  {journal} {\bibinfo  {journal}
  {Physica D}\ }\textbf {\bibinfo {volume} {237}},\ \bibinfo {pages}
  {2856--2865} (\bibinfo {year} {2008})}\BibitemShut {NoStop}%
\bibitem [{\citenamefont {Chugh}\ \emph {et~al.}(2014)\citenamefont {Chugh},
  \citenamefont {Havmoeller}, \citenamefont {Narayanan}, \citenamefont {Singh},
  \citenamefont {Rienstra}, \citenamefont {Benjamin}, \citenamefont {Gillum},
  \citenamefont {Kim}, \citenamefont {McAnulty}, \citenamefont {Zheng},
  \citenamefont {Forouzanfar}, \citenamefont {Ezzati},\ and\ \citenamefont
  {Murray}}]{Chugh_2014}%
  \BibitemOpen
  \bibfield  {author} {\bibinfo {author} {\bibfnamefont {S.~S.}\ \bibnamefont
  {Chugh}}, \bibinfo {author} {\bibfnamefont {R.}~\bibnamefont {Havmoeller}},
  \bibinfo {author} {\bibfnamefont {K.}~\bibnamefont {Narayanan}}, \bibinfo
  {author} {\bibfnamefont {D.}~\bibnamefont {Singh}}, \bibinfo {author}
  {\bibfnamefont {M.}~\bibnamefont {Rienstra}}, \bibinfo {author}
  {\bibfnamefont {E.~J.}\ \bibnamefont {Benjamin}}, \bibinfo {author}
  {\bibfnamefont {R.~F.}\ \bibnamefont {Gillum}}, \bibinfo {author}
  {\bibfnamefont {Y.~H.}\ \bibnamefont {Kim}}, \bibinfo {author} {\bibfnamefont
  {J.~H.~J.}\ \bibnamefont {McAnulty}}, \bibinfo {author} {\bibfnamefont
  {Z.~J.}\ \bibnamefont {Zheng}}, \bibinfo {author} {\bibfnamefont {M.~H.}\
  \bibnamefont {Forouzanfar}}, \bibinfo {author} {\bibfnamefont {M.~N. G. A.
  M.~M.}\ \bibnamefont {Ezzati}}, \ and\ \bibinfo {author} {\bibfnamefont
  {C.~J.}\ \bibnamefont {Murray}},\ }\bibfield  {title} {\enquote {\bibinfo
  {title} {Worldwide epidemiology of atrial fibrillation: a global burden of
  disease 2010 study},}\ }\href@noop {} {\bibfield  {journal} {\bibinfo
  {journal} {Circulation}\ }\textbf {\bibinfo {volume} {129}},\ \bibinfo
  {pages} {837--47} (\bibinfo {year} {2014})}\BibitemShut {NoStop}%
\bibitem [{\citenamefont {Buchwald}, \citenamefont {Norrving},\ and\
  \citenamefont {Petersson}(2016)}]{Buchwald_2016}%
  \BibitemOpen
  \bibfield  {author} {\bibinfo {author} {\bibfnamefont {F.}~\bibnamefont
  {Buchwald}}, \bibinfo {author} {\bibfnamefont {B.}~\bibnamefont {Norrving}},
  \ and\ \bibinfo {author} {\bibfnamefont {J.}~\bibnamefont {Petersson}},\
  }\bibfield  {title} {\enquote {\bibinfo {title} {Atrial fibrillation in
  transient ischemic attack versus ischemic stroke},}\ }\href@noop {}
  {\bibfield  {journal} {\bibinfo  {journal} {Stroke}\ }\textbf {\bibinfo
  {volume} {47}},\ \bibinfo {pages} {2456--2461} (\bibinfo {year}
  {2016})}\BibitemShut {NoStop}%
\bibitem [{\citenamefont {Wolf}, \citenamefont {Abbott},\ and\ \citenamefont
  {Kannel}(1991)}]{Wolf_1991}%
  \BibitemOpen
  \bibfield  {author} {\bibinfo {author} {\bibfnamefont {P.~A.}\ \bibnamefont
  {Wolf}}, \bibinfo {author} {\bibfnamefont {R.~D.}\ \bibnamefont {Abbott}}, \
  and\ \bibinfo {author} {\bibfnamefont {W.~B.}\ \bibnamefont {Kannel}},\
  }\bibfield  {title} {\enquote {\bibinfo {title} {Atrial-fibrillation as an
  independent risk factor for stroke - the framingham-study},}\ }\href@noop {}
  {\bibfield  {journal} {\bibinfo  {journal} {Stroke}\ }\textbf {\bibinfo
  {volume} {22}},\ \bibinfo {pages} {983--988} (\bibinfo {year}
  {1991})}\BibitemShut {NoStop}%
\bibitem [{\citenamefont {Kalantarian}\ \emph {et~al.}(2014)\citenamefont
  {Kalantarian}, \citenamefont {Ay}, \citenamefont {Gollub}, \citenamefont
  {Lee}, \citenamefont {Retzepi}, \citenamefont {Mansour},\ and\ \citenamefont
  {Ruskin}}]{Kalantarian_2014}%
  \BibitemOpen
  \bibfield  {author} {\bibinfo {author} {\bibfnamefont {S.}~\bibnamefont
  {Kalantarian}}, \bibinfo {author} {\bibfnamefont {H.}~\bibnamefont {Ay}},
  \bibinfo {author} {\bibfnamefont {R.~L.}\ \bibnamefont {Gollub}}, \bibinfo
  {author} {\bibfnamefont {H.}~\bibnamefont {Lee}}, \bibinfo {author}
  {\bibfnamefont {K.}~\bibnamefont {Retzepi}}, \bibinfo {author} {\bibfnamefont
  {M.}~\bibnamefont {Mansour}}, \ and\ \bibinfo {author} {\bibfnamefont
  {J.~N.}\ \bibnamefont {Ruskin}},\ }\bibfield  {title} {\enquote {\bibinfo
  {title} {Association between atrial fibrillation and silent cerebral
  infarctions a systematic review and meta-analysis},}\ }\href@noop {}
  {\bibfield  {journal} {\bibinfo  {journal} {Annals of Internal Medicine}\
  }\textbf {\bibinfo {volume} {161}},\ \bibinfo {pages} {650--U158} (\bibinfo
  {year} {2014})}\BibitemShut {NoStop}%
\bibitem [{\citenamefont {Gaita}\ \emph {et~al.}(2013)\citenamefont {Gaita},
  \citenamefont {Corsinovi}, \citenamefont {Anselmino}, \citenamefont
  {Raimondo}, \citenamefont {Pianelli}, \citenamefont {Toso}, \citenamefont
  {Bergamasco}, \citenamefont {Boffano}, \citenamefont {Valentini},
  \citenamefont {Cesarani},\ and\ \citenamefont {Scaglione}}]{Gaita_2013}%
  \BibitemOpen
  \bibfield  {author} {\bibinfo {author} {\bibfnamefont {F.}~\bibnamefont
  {Gaita}}, \bibinfo {author} {\bibfnamefont {L.}~\bibnamefont {Corsinovi}},
  \bibinfo {author} {\bibfnamefont {M.}~\bibnamefont {Anselmino}}, \bibinfo
  {author} {\bibfnamefont {C.}~\bibnamefont {Raimondo}}, \bibinfo {author}
  {\bibfnamefont {M.}~\bibnamefont {Pianelli}}, \bibinfo {author}
  {\bibfnamefont {E.}~\bibnamefont {Toso}}, \bibinfo {author} {\bibfnamefont
  {L.}~\bibnamefont {Bergamasco}}, \bibinfo {author} {\bibfnamefont
  {C.}~\bibnamefont {Boffano}}, \bibinfo {author} {\bibfnamefont {M.~C.}\
  \bibnamefont {Valentini}}, \bibinfo {author} {\bibfnamefont {F.}~\bibnamefont
  {Cesarani}}, \ and\ \bibinfo {author} {\bibfnamefont {M.}~\bibnamefont
  {Scaglione}},\ }\bibfield  {title} {\enquote {\bibinfo {title} {Prevalence of
  silent cerebral ischemia in paroxysmal and persistent atrial fibrillation and
  correlation with cognitive function},}\ }\href@noop {} {\bibfield  {journal}
  {\bibinfo  {journal} {Journal of the American College of Cardiology}\
  }\textbf {\bibinfo {volume} {62}},\ \bibinfo {pages} {1990--1997} (\bibinfo
  {year} {2013})}\BibitemShut {NoStop}%
\bibitem [{\citenamefont {Sabatini}\ \emph {et~al.}(2000)\citenamefont
  {Sabatini}, \citenamefont {Frisoni}, \citenamefont {Barbisoni}, \citenamefont
  {Bellelli}, \citenamefont {Rozzini},\ and\ \citenamefont
  {Trabucchi}}]{Sabatini_2000}%
  \BibitemOpen
  \bibfield  {author} {\bibinfo {author} {\bibfnamefont {T.}~\bibnamefont
  {Sabatini}}, \bibinfo {author} {\bibfnamefont {G.~B.}\ \bibnamefont
  {Frisoni}}, \bibinfo {author} {\bibfnamefont {P.}~\bibnamefont {Barbisoni}},
  \bibinfo {author} {\bibfnamefont {G.}~\bibnamefont {Bellelli}}, \bibinfo
  {author} {\bibfnamefont {R.}~\bibnamefont {Rozzini}}, \ and\ \bibinfo
  {author} {\bibfnamefont {M.}~\bibnamefont {Trabucchi}},\ }\bibfield  {title}
  {\enquote {\bibinfo {title} {Atrial fibrillation and cognitive disorders in
  older people},}\ }\href@noop {} {\bibfield  {journal} {\bibinfo  {journal}
  {Journal of the American Geriatrics Society}\ }\textbf {\bibinfo {volume}
  {48}},\ \bibinfo {pages} {387--390} (\bibinfo {year} {2000})}\BibitemShut
  {NoStop}%
\bibitem [{\citenamefont {Jacobs}\ \emph {et~al.}(2014)\citenamefont {Jacobs},
  \citenamefont {Cutler}, \citenamefont {Day},\ and\ \citenamefont
  {Bunch}}]{Jacobs_2014}%
  \BibitemOpen
  \bibfield  {author} {\bibinfo {author} {\bibfnamefont {V.}~\bibnamefont
  {Jacobs}}, \bibinfo {author} {\bibfnamefont {M.~J.}\ \bibnamefont {Cutler}},
  \bibinfo {author} {\bibfnamefont {J.~D.}\ \bibnamefont {Day}}, \ and\
  \bibinfo {author} {\bibfnamefont {T.~J.}\ \bibnamefont {Bunch}},\ }\bibfield
  {title} {\enquote {\bibinfo {title} {Atrial fibrillation and dementia},}\
  }\href@noop {} {\bibfield  {journal} {\bibinfo  {journal} {Trends in
  Cardiovascular Medicine}\ }\textbf {\bibinfo {volume} {25}},\ \bibinfo
  {pages} {44–--51} (\bibinfo {year} {2014})}\BibitemShut {NoStop}%
\bibitem [{\citenamefont {Kalantarian}\ \emph {et~al.}(2013)\citenamefont
  {Kalantarian}, \citenamefont {Stern}, \citenamefont {Mansour},\ and\
  \citenamefont {Ruskin}}]{Kalantarian_2013}%
  \BibitemOpen
  \bibfield  {author} {\bibinfo {author} {\bibfnamefont {S.}~\bibnamefont
  {Kalantarian}}, \bibinfo {author} {\bibfnamefont {T.~A.}\ \bibnamefont
  {Stern}}, \bibinfo {author} {\bibfnamefont {M.}~\bibnamefont {Mansour}}, \
  and\ \bibinfo {author} {\bibfnamefont {J.~N.}\ \bibnamefont {Ruskin}},\
  }\bibfield  {title} {\enquote {\bibinfo {title} {Cognitive impairment
  associated with atrial fibrillation a meta-analysis},}\ }\href@noop {}
  {\bibfield  {journal} {\bibinfo  {journal} {Annals of Internal Medicine}\
  }\textbf {\bibinfo {volume} {158}},\ \bibinfo {pages} {338--46} (\bibinfo
  {year} {2013})}\BibitemShut {NoStop}%
\bibitem [{\citenamefont {Hui}\ \emph {et~al.}(2015)\citenamefont {Hui},
  \citenamefont {Morley}, \citenamefont {Mikolajczak},\ and\ \citenamefont
  {Lee}}]{Hui_2015}%
  \BibitemOpen
  \bibfield  {author} {\bibinfo {author} {\bibfnamefont {D.~S.}\ \bibnamefont
  {Hui}}, \bibinfo {author} {\bibfnamefont {J.~E.}\ \bibnamefont {Morley}},
  \bibinfo {author} {\bibfnamefont {P.~C.}\ \bibnamefont {Mikolajczak}}, \ and\
  \bibinfo {author} {\bibfnamefont {R.}~\bibnamefont {Lee}},\ }\bibfield
  {title} {\enquote {\bibinfo {title} {Atrial fibrillation: A major risk factor
  for cognitive decline},}\ }\href@noop {} {\bibfield  {journal} {\bibinfo
  {journal} {American Heart Journal}\ }\textbf {\bibinfo {volume} {169}},\
  \bibinfo {pages} {448–--456} (\bibinfo {year} {2015})}\BibitemShut
  {NoStop}%
\bibitem [{\citenamefont {Chen}\ \emph {et~al.}(2016)\citenamefont {Chen},
  \citenamefont {Agarwal}, \citenamefont {Norby}, \citenamefont {Gottesman},
  \citenamefont {Loehr}, \citenamefont {Soliman}, \citenamefont {Mosley},
  \citenamefont {Folsom}, \citenamefont {Coresh},\ and\ \citenamefont
  {Alonso}}]{Chen_2016}%
  \BibitemOpen
  \bibfield  {author} {\bibinfo {author} {\bibfnamefont {L.~Y.}\ \bibnamefont
  {Chen}}, \bibinfo {author} {\bibfnamefont {S.~K.}\ \bibnamefont {Agarwal}},
  \bibinfo {author} {\bibfnamefont {F.~L.}\ \bibnamefont {Norby}}, \bibinfo
  {author} {\bibfnamefont {R.~F.}\ \bibnamefont {Gottesman}}, \bibinfo {author}
  {\bibfnamefont {L.~R.}\ \bibnamefont {Loehr}}, \bibinfo {author}
  {\bibfnamefont {E.~Z.}\ \bibnamefont {Soliman}}, \bibinfo {author}
  {\bibfnamefont {T.~H.}\ \bibnamefont {Mosley}}, \bibinfo {author}
  {\bibfnamefont {A.~R.}\ \bibnamefont {Folsom}}, \bibinfo {author}
  {\bibfnamefont {J.}~\bibnamefont {Coresh}}, \ and\ \bibinfo {author}
  {\bibfnamefont {A.}~\bibnamefont {Alonso}},\ }\bibfield  {title} {\enquote
  {\bibinfo {title} {Persistent but not paroxysmal atrial fibrillation is
  independently associated with lower cognitive function},}\ }\href@noop {}
  {\bibfield  {journal} {\bibinfo  {journal} {Journal of the American College
  of Cardiology}\ }\textbf {\bibinfo {volume} {67}},\ \bibinfo {pages}
  {1379--1380} (\bibinfo {year} {2016})}\BibitemShut {NoStop}%
\bibitem [{\citenamefont {Thacker}\ \emph {et~al.}(2013)\citenamefont
  {Thacker}, \citenamefont {McKnight}, \citenamefont {Psaty}, \citenamefont
  {Longstreth}, \citenamefont {Sitlani}, \citenamefont {Dublin}, \citenamefont
  {Arnold}, \citenamefont {Fitzpatrick}, \citenamefont {Gottesman},\ and\
  \citenamefont {Heckbert}}]{Thacker_2013}%
  \BibitemOpen
  \bibfield  {author} {\bibinfo {author} {\bibfnamefont {E.~L.}\ \bibnamefont
  {Thacker}}, \bibinfo {author} {\bibfnamefont {B.}~\bibnamefont {McKnight}},
  \bibinfo {author} {\bibfnamefont {B.~M.}\ \bibnamefont {Psaty}}, \bibinfo
  {author} {\bibfnamefont {W.~T.~J.}\ \bibnamefont {Longstreth}}, \bibinfo
  {author} {\bibfnamefont {C.~M.}\ \bibnamefont {Sitlani}}, \bibinfo {author}
  {\bibfnamefont {S.}~\bibnamefont {Dublin}}, \bibinfo {author} {\bibfnamefont
  {A.~M.}\ \bibnamefont {Arnold}}, \bibinfo {author} {\bibfnamefont {A.~L.}\
  \bibnamefont {Fitzpatrick}}, \bibinfo {author} {\bibfnamefont {R.~F.}\
  \bibnamefont {Gottesman}}, \ and\ \bibinfo {author} {\bibfnamefont {S.~R.}\
  \bibnamefont {Heckbert}},\ }\bibfield  {title} {\enquote {\bibinfo {title}
  {Atrial fibrillation and cognitive decline a longitudinal cohort study},}\
  }\href@noop {} {\bibfield  {journal} {\bibinfo  {journal} {Neurology}\
  }\textbf {\bibinfo {volume} {81}},\ \bibinfo {pages} {119--125} (\bibinfo
  {year} {2013})}\BibitemShut {NoStop}%
\bibitem [{\citenamefont {Kanmanthareddy}\ \emph {et~al.}(2014)\citenamefont
  {Kanmanthareddy}, \citenamefont {Vallakati}, \citenamefont {Sridhar},
  \citenamefont {Reddy}, \citenamefont {Sanjani}, \citenamefont {Pillarisetti},
  \citenamefont {Atkins}, \citenamefont {Bommana}, \citenamefont {Jaeger},
  \citenamefont {Berenbom},\ and\ \citenamefont {Lakkireddy}}]{Kanmanthareddy}%
  \BibitemOpen
  \bibfield  {author} {\bibinfo {author} {\bibfnamefont {A.}~\bibnamefont
  {Kanmanthareddy}}, \bibinfo {author} {\bibfnamefont {A.}~\bibnamefont
  {Vallakati}}, \bibinfo {author} {\bibfnamefont {A.}~\bibnamefont {Sridhar}},
  \bibinfo {author} {\bibfnamefont {M.}~\bibnamefont {Reddy}}, \bibinfo
  {author} {\bibfnamefont {H.~P.}\ \bibnamefont {Sanjani}}, \bibinfo {author}
  {\bibfnamefont {J.}~\bibnamefont {Pillarisetti}}, \bibinfo {author}
  {\bibfnamefont {D.}~\bibnamefont {Atkins}}, \bibinfo {author} {\bibfnamefont
  {S.}~\bibnamefont {Bommana}}, \bibinfo {author} {\bibfnamefont
  {M.}~\bibnamefont {Jaeger}}, \bibinfo {author} {\bibfnamefont
  {L.}~\bibnamefont {Berenbom}}, \ and\ \bibinfo {author} {\bibfnamefont
  {D.}~\bibnamefont {Lakkireddy}},\ }\bibfield  {title} {\enquote {\bibinfo
  {title} {The impact of atrial fibrillation and its treatment on dementia},}\
  }\href@noop {} {\bibfield  {journal} {\bibinfo  {journal} {Current Cardiology
  Reports}\ }\textbf {\bibinfo {volume} {16}},\ \bibinfo {pages} {519}
  (\bibinfo {year} {2014})}\BibitemShut {NoStop}%
\bibitem [{\citenamefont {Severi}, \citenamefont {Rodriguez},\ and\
  \citenamefont {Zaza}(2014)}]{Severi}%
  \BibitemOpen
  \bibfield  {author} {\bibinfo {author} {\bibfnamefont {S.}~\bibnamefont
  {Severi}}, \bibinfo {author} {\bibfnamefont {B.}~\bibnamefont {Rodriguez}}, \
  and\ \bibinfo {author} {\bibfnamefont {A.}~\bibnamefont {Zaza}},\ }\bibfield
  {title} {\enquote {\bibinfo {title} {Computational cardiac electrophysiology
  is ready for prime time},}\ }\href@noop {} {\bibfield  {journal} {\bibinfo
  {journal} {Europace}\ }\textbf {\bibinfo {volume} {16}},\ \bibinfo {pages}
  {382–--3} (\bibinfo {year} {2014})}\BibitemShut {NoStop}%
\bibitem [{\citenamefont {Shi}, \citenamefont {Lawford},\ and\ \citenamefont
  {Hose}(2011)}]{Shi}%
  \BibitemOpen
  \bibfield  {author} {\bibinfo {author} {\bibfnamefont {Y.}~\bibnamefont
  {Shi}}, \bibinfo {author} {\bibfnamefont {P.}~\bibnamefont {Lawford}}, \ and\
  \bibinfo {author} {\bibfnamefont {R.}~\bibnamefont {Hose}},\ }\bibfield
  {title} {\enquote {\bibinfo {title} {Review of zero-d and 1-d models of blood
  flow in the cardiovascular system},}\ }\href@noop {} {\bibfield  {journal}
  {\bibinfo  {journal} {BioMedical Engineering OnLine}\ }\textbf {\bibinfo
  {volume} {10}},\ \bibinfo {pages} {33} (\bibinfo {year} {2011})}\BibitemShut
  {NoStop}%
\bibitem [{\citenamefont {Anselmino}\ \emph {et~al.}(2016)\citenamefont
  {Anselmino}, \citenamefont {Scarsoglio}, \citenamefont {Saglietto},
  \citenamefont {Gaita},\ and\ \citenamefont {Ridolfi}}]{SR_2016}%
  \BibitemOpen
  \bibfield  {author} {\bibinfo {author} {\bibfnamefont {M.}~\bibnamefont
  {Anselmino}}, \bibinfo {author} {\bibfnamefont {S.}~\bibnamefont
  {Scarsoglio}}, \bibinfo {author} {\bibfnamefont {A.}~\bibnamefont
  {Saglietto}}, \bibinfo {author} {\bibfnamefont {F.}~\bibnamefont {Gaita}}, \
  and\ \bibinfo {author} {\bibfnamefont {L.}~\bibnamefont {Ridolfi}},\
  }\bibfield  {title} {\enquote {\bibinfo {title} {Transient cerebral
  hypoperfusion and hypertensive events during atrial fibrillation: a plausible
  mechanism for cognitive impairment},}\ }\href@noop {} {\bibfield  {journal}
  {\bibinfo  {journal} {Scientific Reports}\ }\textbf {\bibinfo {volume} {6}},\
  \bibinfo {pages} {28635} (\bibinfo {year} {2016})}\BibitemShut {NoStop}%
\bibitem [{\citenamefont {Scarsoglio}\ \emph {et~al.}(2017)\citenamefont
  {Scarsoglio}, \citenamefont {Saglietto}, \citenamefont {Anselmino},
  \citenamefont {Gaita},\ and\ \citenamefont {Ridolfi}}]{Scarsoglio_JRSI}%
  \BibitemOpen
  \bibfield  {author} {\bibinfo {author} {\bibfnamefont {S.}~\bibnamefont
  {Scarsoglio}}, \bibinfo {author} {\bibfnamefont {A.}~\bibnamefont
  {Saglietto}}, \bibinfo {author} {\bibfnamefont {M.}~\bibnamefont
  {Anselmino}}, \bibinfo {author} {\bibfnamefont {F.}~\bibnamefont {Gaita}}, \
  and\ \bibinfo {author} {\bibfnamefont {L.}~\bibnamefont {Ridolfi}},\
  }\bibfield  {title} {\enquote {\bibinfo {title} {Alteration of
  cerebrovascular haemodynamic patterns due to atrial fibrillation: an in
  silico investigation},}\ }\href@noop {} {\bibfield  {journal} {\bibinfo
  {journal} {Journal of The Royal Society Interface}\ }\textbf {\bibinfo
  {volume} {14}},\ \bibinfo {pages} {20170180} (\bibinfo {year}
  {2017})}\BibitemShut {NoStop}%
\bibitem [{\citenamefont {Korakianitis}\ and\ \citenamefont
  {Shi}(2006)}]{Korakianitis}%
  \BibitemOpen
  \bibfield  {author} {\bibinfo {author} {\bibfnamefont {T.}~\bibnamefont
  {Korakianitis}}\ and\ \bibinfo {author} {\bibfnamefont {Y.}~\bibnamefont
  {Shi}},\ }\bibfield  {title} {\enquote {\bibinfo {title} {Numerical
  simulation of cardiovascular dynamics with healthy and diseased heart
  valves},}\ }\href@noop {} {\bibfield  {journal} {\bibinfo  {journal} {Journal
  of Biomechanics}\ }\textbf {\bibinfo {volume} {39}},\ \bibinfo {pages}
  {1964–--1982} (\bibinfo {year} {2006})}\BibitemShut {NoStop}%
\bibitem [{\citenamefont {Scarsoglio}\ \emph {et~al.}(2014)\citenamefont
  {Scarsoglio}, \citenamefont {Guala}, \citenamefont {Camporeale},\ and\
  \citenamefont {Ridolfi}}]{MBEC_2014}%
  \BibitemOpen
  \bibfield  {author} {\bibinfo {author} {\bibfnamefont {S.}~\bibnamefont
  {Scarsoglio}}, \bibinfo {author} {\bibfnamefont {A.}~\bibnamefont {Guala}},
  \bibinfo {author} {\bibfnamefont {C.}~\bibnamefont {Camporeale}}, \ and\
  \bibinfo {author} {\bibfnamefont {L.}~\bibnamefont {Ridolfi}},\ }\bibfield
  {title} {\enquote {\bibinfo {title} {Impact of atrial fibrillation on the
  cardiovascular system through a lumped-parameter approach},}\ }\href@noop {}
  {\bibfield  {journal} {\bibinfo  {journal} {Medical and Biological
  Engineering and Computing}\ }\textbf {\bibinfo {volume} {52}},\ \bibinfo
  {pages} {905--920} (\bibinfo {year} {2014})}\BibitemShut {NoStop}%
\bibitem [{\citenamefont {Scarsoglio}\ \emph
  {et~al.}(2016{\natexlab{a}})\citenamefont {Scarsoglio}, \citenamefont
  {Camporeale}, \citenamefont {Guala},\ and\ \citenamefont
  {Ridolfi}}]{CMBBE_2016}%
  \BibitemOpen
  \bibfield  {author} {\bibinfo {author} {\bibfnamefont {S.}~\bibnamefont
  {Scarsoglio}}, \bibinfo {author} {\bibfnamefont {C.}~\bibnamefont
  {Camporeale}}, \bibinfo {author} {\bibfnamefont {A.}~\bibnamefont {Guala}}, \
  and\ \bibinfo {author} {\bibfnamefont {L.}~\bibnamefont {Ridolfi}},\
  }\bibfield  {title} {\enquote {\bibinfo {title} {Fluid dynamics of heart
  valves during atrial fibrillation: a lumped parameter-based approach},}\
  }\href@noop {} {\bibfield  {journal} {\bibinfo  {journal} {Computer Methods
  in Biomechanics and Biomedical Engineering}\ }\textbf {\bibinfo {volume}
  {10}},\ \bibinfo {pages} {1060–--1068} (\bibinfo {year}
  {2016}{\natexlab{a}})}\BibitemShut {NoStop}%
\bibitem [{\citenamefont {Ursino}\ and\ \citenamefont
  {Giannessi}(2010)}]{Ursino_2010}%
  \BibitemOpen
  \bibfield  {author} {\bibinfo {author} {\bibfnamefont {M.}~\bibnamefont
  {Ursino}}\ and\ \bibinfo {author} {\bibfnamefont {M.}~\bibnamefont
  {Giannessi}},\ }\bibfield  {title} {\enquote {\bibinfo {title} {A model of
  cerebrovascular reactivity including the circle of willis and cortical
  anastomoses},}\ }\href@noop {} {\bibfield  {journal} {\bibinfo  {journal}
  {Annals of Biomedical Engineering}\ }\textbf {\bibinfo {volume} {38}},\
  \bibinfo {pages} {955–--974} (\bibinfo {year} {2010})}\BibitemShut
  {NoStop}%
\bibitem [{\citenamefont {Hayano}\ \emph {et~al.}(1997)\citenamefont {Hayano},
  \citenamefont {Yamasaki}, \citenamefont {Sakata}, \citenamefont {Okada},
  \citenamefont {Mukai},\ and\ \citenamefont {Fujinami}}]{Hayano_1997}%
  \BibitemOpen
  \bibfield  {author} {\bibinfo {author} {\bibfnamefont {J.}~\bibnamefont
  {Hayano}}, \bibinfo {author} {\bibfnamefont {F.}~\bibnamefont {Yamasaki}},
  \bibinfo {author} {\bibfnamefont {S.}~\bibnamefont {Sakata}}, \bibinfo
  {author} {\bibfnamefont {A.}~\bibnamefont {Okada}}, \bibinfo {author}
  {\bibfnamefont {S.}~\bibnamefont {Mukai}}, \ and\ \bibinfo {author}
  {\bibfnamefont {T.}~\bibnamefont {Fujinami}},\ }\bibfield  {title} {\enquote
  {\bibinfo {title} {Spectral characteristics of ventricular response to atrial
  fibrillation},}\ }\href@noop {} {\bibfield  {journal} {\bibinfo  {journal}
  {American Journal of Physiology- Heart and Circulatory Physiology}\ }\textbf
  {\bibinfo {volume} {273}},\ \bibinfo {pages} {H2811--H2816} (\bibinfo {year}
  {1997})}\BibitemShut {NoStop}%
\bibitem [{\citenamefont {Hennig}\ \emph {et~al.}(2006)\citenamefont {Hennig},
  \citenamefont {Maass}, \citenamefont {Hayano},\ and\ \citenamefont
  {Heinrichs}}]{Hennig_2006}%
  \BibitemOpen
  \bibfield  {author} {\bibinfo {author} {\bibfnamefont {T.}~\bibnamefont
  {Hennig}}, \bibinfo {author} {\bibfnamefont {P.}~\bibnamefont {Maass}},
  \bibinfo {author} {\bibfnamefont {J.}~\bibnamefont {Hayano}}, \ and\ \bibinfo
  {author} {\bibfnamefont {S.}~\bibnamefont {Heinrichs}},\ }\bibfield  {title}
  {\enquote {\bibinfo {title} {Exponential distribution of long heart beat
  intervals during atrial fibrillation and their relevance for white noise
  behaviour in power spectrum},}\ }\href@noop {} {\bibfield  {journal}
  {\bibinfo  {journal} {Journal of Biological Physics}\ }\textbf {\bibinfo
  {volume} {32}},\ \bibinfo {pages} {383–--392} (\bibinfo {year}
  {2006})}\BibitemShut {NoStop}%
\bibitem [{\citenamefont {Kobayashi}\ and\ \citenamefont
  {Musha}(1982)}]{Musha1982}%
  \BibitemOpen
  \bibfield  {author} {\bibinfo {author} {\bibfnamefont {M.}~\bibnamefont
  {Kobayashi}}\ and\ \bibinfo {author} {\bibfnamefont {T.}~\bibnamefont
  {Musha}},\ }\bibfield  {title} {\enquote {\bibinfo {title} {1/f fluctuation
  of heartbeat period},}\ }\href@noop {} {\bibfield  {journal} {\bibinfo
  {journal} {IEEE Transactions on Biomedical Engineering}\ }\textbf {\bibinfo
  {volume} {29}},\ \bibinfo {pages} {456--457} (\bibinfo {year}
  {1982})}\BibitemShut {NoStop}%
\bibitem [{\citenamefont {Peng}\ \emph {et~al.}(1993)\citenamefont {Peng},
  \citenamefont {Mietus}, \citenamefont {Hausdorff}, \citenamefont {Havlin},
  \citenamefont {Stanley},\ and\ \citenamefont {Goldberger}}]{Peng1993}%
  \BibitemOpen
  \bibfield  {author} {\bibinfo {author} {\bibfnamefont {C.~K.}\ \bibnamefont
  {Peng}}, \bibinfo {author} {\bibfnamefont {J.}~\bibnamefont {Mietus}},
  \bibinfo {author} {\bibfnamefont {J.~M.}\ \bibnamefont {Hausdorff}}, \bibinfo
  {author} {\bibfnamefont {S.}~\bibnamefont {Havlin}}, \bibinfo {author}
  {\bibfnamefont {H.~E.}\ \bibnamefont {Stanley}}, \ and\ \bibinfo {author}
  {\bibfnamefont {A.~L.}\ \bibnamefont {Goldberger}},\ }\bibfield  {title}
  {\enquote {\bibinfo {title} {Long-range anticorrelations and non-gaussian
  behavior of the heartbeat},}\ }\href@noop {} {\bibfield  {journal} {\bibinfo
  {journal} {Physical Review Letters}\ }\textbf {\bibinfo {volume} {70}},\
  \bibinfo {pages} {1343--6} (\bibinfo {year} {1993})}\BibitemShut {NoStop}%
\bibitem [{\citenamefont {Saul}\ \emph {et~al.}(1987)\citenamefont {Saul},
  \citenamefont {Albrecht}, \citenamefont {Berger},\ and\ \citenamefont
  {Cohen}}]{Saul1987}%
  \BibitemOpen
  \bibfield  {author} {\bibinfo {author} {\bibfnamefont {J.~P.}\ \bibnamefont
  {Saul}}, \bibinfo {author} {\bibfnamefont {P.}~\bibnamefont {Albrecht}},
  \bibinfo {author} {\bibfnamefont {R.~D.}\ \bibnamefont {Berger}}, \ and\
  \bibinfo {author} {\bibfnamefont {R.~J.}\ \bibnamefont {Cohen}},\ }\bibfield
  {title} {\enquote {\bibinfo {title} {Analysis of long term heart rate
  variability: methods, 1/f scaling and implications},}\ }\href@noop {}
  {\bibfield  {journal} {\bibinfo  {journal} {Computers in Cardiology}\
  }\textbf {\bibinfo {volume} {14}},\ \bibinfo {pages} {419--422} (\bibinfo
  {year} {1987})}\BibitemShut {NoStop}%
\bibitem [{\citenamefont {Yamamoto}\ and\ \citenamefont
  {Hughson}(1994)}]{Yamamoto1994}%
  \BibitemOpen
  \bibfield  {author} {\bibinfo {author} {\bibfnamefont {Y.}~\bibnamefont
  {Yamamoto}}\ and\ \bibinfo {author} {\bibfnamefont {R.~L.}\ \bibnamefont
  {Hughson}},\ }\bibfield  {title} {\enquote {\bibinfo {title} {On the fractal
  nature of heart rate variability in humans: effects of data length and
  beta-adrenergic blockade},}\ }\href@noop {} {\bibfield  {journal} {\bibinfo
  {journal} {American Journal of Physiology}\ }\textbf {\bibinfo {volume}
  {266}},\ \bibinfo {pages} {R40} (\bibinfo {year} {1994})}\BibitemShut
  {NoStop}%
\bibitem [{\citenamefont {Bootsma}\ \emph {et~al.}(1970)\citenamefont
  {Bootsma}, \citenamefont {Hoelen}, \citenamefont {Strackee},\ and\
  \citenamefont {Meijler}}]{Bootsma}%
  \BibitemOpen
  \bibfield  {author} {\bibinfo {author} {\bibfnamefont {B.~K.}\ \bibnamefont
  {Bootsma}}, \bibinfo {author} {\bibfnamefont {A.~J.}\ \bibnamefont {Hoelen}},
  \bibinfo {author} {\bibfnamefont {J.}~\bibnamefont {Strackee}}, \ and\
  \bibinfo {author} {\bibfnamefont {F.~L.}\ \bibnamefont {Meijler}},\
  }\bibfield  {title} {\enquote {\bibinfo {title} {Analysis of r-r intervals in
  patients with atrial fibrillation at rest and during exercise},}\ }\href@noop
  {} {\bibfield  {journal} {\bibinfo  {journal} {Circulation}\ }\textbf
  {\bibinfo {volume} {41}},\ \bibinfo {pages} {783–--794} (\bibinfo {year}
  {1970})}\BibitemShut {NoStop}%
\bibitem [{\citenamefont {Sosnowski}\ \emph {et~al.}(2011)\citenamefont
  {Sosnowski}, \citenamefont {Korzeniowska}, \citenamefont {Macfarlane},\ and\
  \citenamefont {Tendera}}]{Sosnowski}%
  \BibitemOpen
  \bibfield  {author} {\bibinfo {author} {\bibfnamefont {M.}~\bibnamefont
  {Sosnowski}}, \bibinfo {author} {\bibfnamefont {B.}~\bibnamefont
  {Korzeniowska}}, \bibinfo {author} {\bibfnamefont {P.}~\bibnamefont
  {Macfarlane}}, \ and\ \bibinfo {author} {\bibfnamefont {M.}~\bibnamefont
  {Tendera}},\ }\bibfield  {title} {\enquote {\bibinfo {title} {Relationship
  between r–r interval variation and left ventricular function in sinus
  rhythm and atrial fibrillation as estimated by means of heart rate
  variability fraction},}\ }\href@noop {} {\bibfield  {journal} {\bibinfo
  {journal} {Cardiology Journal}\ }\textbf {\bibinfo {volume} {18}},\ \bibinfo
  {pages} {538–--545} (\bibinfo {year} {2011})}\BibitemShut {NoStop}%
\bibitem [{\citenamefont {Tateno}\ and\ \citenamefont {Glass}(2001)}]{Tateno}%
  \BibitemOpen
  \bibfield  {author} {\bibinfo {author} {\bibfnamefont {K.}~\bibnamefont
  {Tateno}}\ and\ \bibinfo {author} {\bibfnamefont {L.}~\bibnamefont {Glass}},\
  }\bibfield  {title} {\enquote {\bibinfo {title} {Automatic detection of
  atrial fibrillation using the coefficient of variation and density histograms
  of rr and $\delta$ rr intervals},}\ }\href@noop {} {\bibfield  {journal}
  {\bibinfo  {journal} {Medical and Biological Engineering and Computing}\
  }\textbf {\bibinfo {volume} {39}},\ \bibinfo {pages} {664–--671} (\bibinfo
  {year} {2001})}\BibitemShut {NoStop}%
\bibitem [{\citenamefont {Westerhof}, \citenamefont {Stergiopulos},\ and\
  \citenamefont {Noble}(2010)}]{westerhof1}%
  \BibitemOpen
  \bibfield  {author} {\bibinfo {author} {\bibfnamefont {N.}~\bibnamefont
  {Westerhof}}, \bibinfo {author} {\bibfnamefont {N.}~\bibnamefont
  {Stergiopulos}}, \ and\ \bibinfo {author} {\bibfnamefont {M.~I.}\
  \bibnamefont {Noble}},\ }\href@noop {} {\emph {\bibinfo {title} {Snapshots of
  Hemodynamics: An aid for Clinical Research and Graduate Education}}}\
  (\bibinfo  {publisher} {Springer, New York},\ \bibinfo {year}
  {2010})\BibitemShut {NoStop}%
\bibitem [{\citenamefont {Scarsoglio}\ \emph
  {et~al.}(2016{\natexlab{b}})\citenamefont {Scarsoglio}, \citenamefont
  {Saglietto}, \citenamefont {Gaita}, \citenamefont {Ridolfi},\ and\
  \citenamefont {Anselmino}}]{PeerJ_2016}%
  \BibitemOpen
  \bibfield  {author} {\bibinfo {author} {\bibfnamefont {S.}~\bibnamefont
  {Scarsoglio}}, \bibinfo {author} {\bibfnamefont {A.}~\bibnamefont
  {Saglietto}}, \bibinfo {author} {\bibfnamefont {F.}~\bibnamefont {Gaita}},
  \bibinfo {author} {\bibfnamefont {L.}~\bibnamefont {Ridolfi}}, \ and\
  \bibinfo {author} {\bibfnamefont {M.}~\bibnamefont {Anselmino}},\ }\bibfield
  {title} {\enquote {\bibinfo {title} {Computational fluid dynamics modelling
  of left valvular heart diseases during atrial fibrillation},}\ }\href@noop {}
  {\bibfield  {journal} {\bibinfo  {journal} {PeerJ}\ }\textbf {\bibinfo
  {volume} {4}},\ \bibinfo {pages} {e2240} (\bibinfo {year}
  {2016}{\natexlab{b}})}\BibitemShut {NoStop}%
\bibitem [{\citenamefont {Anselmino}\ \emph {et~al.}(2015)\citenamefont
  {Anselmino}, \citenamefont {Scarsoglio}, \citenamefont {Camporeale},
  \citenamefont {Saglietto}, \citenamefont {Gaita},\ and\ \citenamefont
  {Ridolfi}}]{PlosOne_2015}%
  \BibitemOpen
  \bibfield  {author} {\bibinfo {author} {\bibfnamefont {M.}~\bibnamefont
  {Anselmino}}, \bibinfo {author} {\bibfnamefont {S.}~\bibnamefont
  {Scarsoglio}}, \bibinfo {author} {\bibfnamefont {C.}~\bibnamefont
  {Camporeale}}, \bibinfo {author} {\bibfnamefont {A.}~\bibnamefont
  {Saglietto}}, \bibinfo {author} {\bibfnamefont {F.}~\bibnamefont {Gaita}}, \
  and\ \bibinfo {author} {\bibfnamefont {L.}~\bibnamefont {Ridolfi}},\
  }\bibfield  {title} {\enquote {\bibinfo {title} {Rate control management of
  atrial fibrillation: may a mathematical model suggest an ideal heart rate?}}\
  }\href@noop {} {\bibfield  {journal} {\bibinfo  {journal} {PLoS ONE}\
  }\textbf {\bibinfo {volume} {10}},\ \bibinfo {pages} {e119868} (\bibinfo
  {year} {2015})}\BibitemShut {NoStop}%
\bibitem [{\citenamefont {Anselmino}\ \emph {et~al.}(2017)\citenamefont
  {Anselmino}, \citenamefont {Scarsoglio}, \citenamefont {Saglietto},
  \citenamefont {Gaita},\ and\ \citenamefont {Ridolfi}}]{PlosOne_2017}%
  \BibitemOpen
  \bibfield  {author} {\bibinfo {author} {\bibfnamefont {M.}~\bibnamefont
  {Anselmino}}, \bibinfo {author} {\bibfnamefont {S.}~\bibnamefont
  {Scarsoglio}}, \bibinfo {author} {\bibfnamefont {A.}~\bibnamefont
  {Saglietto}}, \bibinfo {author} {\bibfnamefont {F.}~\bibnamefont {Gaita}}, \
  and\ \bibinfo {author} {\bibfnamefont {L.}~\bibnamefont {Ridolfi}},\
  }\bibfield  {title} {\enquote {\bibinfo {title} {A computational study on the
  relation between resting heart rate and atrial fibrillation hemodynamics
  under exercise},}\ }\href@noop {} {\bibfield  {journal} {\bibinfo  {journal}
  {PLoS ONE}\ }\textbf {\bibinfo {volume} {12}},\ \bibinfo {pages} {e0169967}
  (\bibinfo {year} {2017})}\BibitemShut {NoStop}%
\bibitem [{\citenamefont {Newman}(2010)}]{newman2010}%
  \BibitemOpen
  \bibfield  {author} {\bibinfo {author} {\bibfnamefont {M.~E.~J.}\
  \bibnamefont {Newman}},\ }\href@noop {} {\emph {\bibinfo {title} {Networks:
  An Introduction}}}\ (\bibinfo  {publisher} {Oxford University Press},\
  \bibinfo {year} {2010})\BibitemShut {NoStop}%
\bibitem [{\citenamefont {Newman}(2002)}]{Newman_2002}%
  \BibitemOpen
  \bibfield  {author} {\bibinfo {author} {\bibfnamefont {M.~E.~J.}\
  \bibnamefont {Newman}},\ }\bibfield  {title} {\enquote {\bibinfo {title}
  {Assortative mixing in networks},}\ }\href@noop {} {\bibfield  {journal}
  {\bibinfo  {journal} {Physical Review Letters}\ }\textbf {\bibinfo {volume}
  {89}},\ \bibinfo {pages} {208701} (\bibinfo {year} {2002})}\BibitemShut
  {NoStop}%
\bibitem [{\citenamefont {Valente}\ \emph {et~al.}(2008)\citenamefont
  {Valente}, \citenamefont {Coronges}, \citenamefont {Lakon},\ and\
  \citenamefont {Costenbader}}]{Valente}%
  \BibitemOpen
  \bibfield  {author} {\bibinfo {author} {\bibfnamefont {T.~W.}\ \bibnamefont
  {Valente}}, \bibinfo {author} {\bibfnamefont {K.}~\bibnamefont {Coronges}},
  \bibinfo {author} {\bibfnamefont {C.}~\bibnamefont {Lakon}}, \ and\ \bibinfo
  {author} {\bibfnamefont {E.}~\bibnamefont {Costenbader}},\ }\bibfield
  {title} {\enquote {\bibinfo {title} {How correlated are network centrality
  measures?}}\ }\href@noop {} {\bibfield  {journal} {\bibinfo  {journal}
  {Connect (Tor)}\ }\textbf {\bibinfo {volume} {28}},\ \bibinfo {pages}
  {16–--26} (\bibinfo {year} {2008})}\BibitemShut {NoStop}%
\bibitem [{\citenamefont {Bastian}, \citenamefont {Heymann},\ and\
  \citenamefont {Jacomy}(2009)}]{gephi}%
  \BibitemOpen
  \bibfield  {author} {\bibinfo {author} {\bibfnamefont {M.}~\bibnamefont
  {Bastian}}, \bibinfo {author} {\bibfnamefont {S.}~\bibnamefont {Heymann}}, \
  and\ \bibinfo {author} {\bibfnamefont {M.}~\bibnamefont {Jacomy}},\
  }\bibfield  {title} {\enquote {\bibinfo {title} {Gephi: An open source
  software for exploring and manipulating networks},}\ }in\ \href@noop {}
  {\emph {\bibinfo {booktitle} {International AAAI Conference on Weblogs and
  Social Media}}}\ (\bibinfo {year} {2009})\BibitemShut {NoStop}%
\end{thebibliography}%

\end{document}